\documentclass[floats,floatfix,amssymb,prd,superscriptaddress,nofootinbib,preprintnumbers]{revtex4-1}

\usepackage{subcaption,ragged2e}
\usepackage{lipsum}
\DeclareCaptionJustification{justified}{\justifying}
\usepackage{amssymb,amsmath,verbatim,mathtools,needspace,enumitem,etoolbox,graphicx,microtype,afterpage,bm}
\usepackage{tikz}
\usepackage{framed}
\usepackage{hyperref}
\hypersetup{colorlinks, citecolor=bluscuro, linkcolor=black, urlcolor=bluscuro}
\definecolor{rossos}{cmyk}{0,1,1,0.55}
\definecolor{bluscuro}{rgb}{0.15, 0.2, .85}
\definecolor{bluchiaro}{cmyk}{1,.3,0.,0.1}
\definecolor{ForestGreen}{rgb}{0.13, 0.55, 0.13}
\definecolor{TLGreen}{RGB}{50, 164, 49}
\definecolor{TLOrange}{RGB}{231,180,22}
\definecolor{TLRed}{RGB}{204,50,50}

\usepackage{multirow,pifont,lmodern,float,tabularx,booktabs}
\usepackage[all]{hypcap}
\usepackage[T1]{fontenc}
\usepackage[utf8]{inputenc}
\usepackage{cleveref}
\usepackage{ulem}

\usepackage{dsfont}

\renewcommand{\arraystretch}{1.4}

\allowdisplaybreaks

\newcommand{\be}{\begin{equation}}
\newcommand{\ee}{\end{equation}}

\newcommand{\Nd}{{N_{\rm det}}}
\newcommand{\Ns}{N_s}
\newcommand{\Nf}{N_f}

\newcommand{\Npairs}{N_{\rm pairs}}
\newcommand{\ML}[1]{{\hat{#1}}}

\newcommand{\xmarkred}{\textcolor{red}{\ding{55}}}
\newcommand{\cmarkgreen}{\textcolor{ForestGreen}{\ding{51}}}

\newcommand{\cern}{
CERN, Theoretical Physics Department,
Esplanade des Particules 1, Geneva 1211, Switzerland}

\newcommand{\iem}{
Instituto de Estructura de la Materia (IEM), CSIC, Serrano 121, 28006 Madrid, Spain}
\newcommand{\unipi}{
Dipartimento di Fisica ``Enrico Fermi'', Universit\`a di Pisa, Largo Bruno Pontecorvo 3, Pisa I-56127, Italy}
\newcommand{\infnpi}{
INFN, Sezione di Pisa, Largo Bruno Pontecorvo 3, Pisa I-56127, Italy}

\begin{document}

\title{Likelihoods for Stochastic Gravitational Wave Background Data Analysis}

\author{Gabriele Franciolini}
\email{gabriele.franciolini@cern.ch}
\affiliation{\cern} 

\author{Mauro Pieroni}
\email{mauro.pieroni@csic.es}
\affiliation{\cern}
\affiliation{\iem}

\author{Angelo Ricciardone}
\email{angelo.ricciardone@unipi.it}
\affiliation{\unipi} 
\affiliation{\infnpi} 

\author{Joseph D.~Romano}
\email{joseph.romano@utrgv.edu}
\affiliation{Department of Physics and Astronomy,
University of Texas Rio Grande Valley,
One West University Boulevard,
Brownsville, TX 78520, USA}

\date{\today}

\begin{abstract}
\noindent
We present a systematic study of likelihood functions used for Stochastic Gravitational-Wave Background (SGWB) searches. By dividing the data into many short segments, one customarily takes advantage of the Central Limit Theorem to justify a Gaussian cross-correlation likelihood. We show, with a hierarchy of ever more realistic examples—beginning with a single frequency bin and one detector, and then moving to two and three detectors with white and colored signal and noise—that approximating the exact Whittle likelihood by various Gaussian alternatives can induce systematic biases in the estimation of the SGWB parameters. We derive several approximations for the full likelihood and identify regimes where Gaussianity breaks down. We also discuss the possibility of conditioning the full likelihood on fiducial noise estimates to produce unbiased SGWB parameter estimation. We show that for some segment durations and bandwidths, particularly in space-based and pulsar-timing arrays, the bias can exceed the statistical uncertainty. Our results provide practical guidance for segment choice, likelihood selection, and data-compression strategies to ensure robust SGWB inference in current and next-generation gravitational-wave detectors.
\end{abstract}

\preprint{CERN-TH-2025-094}

\maketitle

\tableofcontents 

\section{Introduction}\label{intro}

The detection of a stochastic gravitational-wave background (SGWB) is a primary goal of current and future GW detectors. Arising from the superposition of unresolved astrophysical and/or cosmological sources, the SGWB provides a unique opportunity to investigate the population properties of different astrophysical compact objects~\cite{Regimbau:2011rp} and/or probe the physics of the early universe~\cite{Caprini:2018mtu}.

Ground-based detectors such as LIGO~\cite{LIGO}, Virgo~\cite{Virgo}, and KAGRA~\cite{KAGRA} (LVK) at their design sensitivity, together with future detectors such as Einstein Telescope (ET)~\cite{ET:2019dnz,Abac:2025saz} and Cosmic Explorer (CE)~\cite{Reitze:2019iox}, and space-based interferometers like LISA~\cite{LISA:2024hlh}, will significantly improve the sensitivity to the SGWB. A key technique for SGWB searches with current ground-based detectors consists of cross-correlating the output of multiple detectors (or channels)~\cite{Allen:1997ad,Romano:2016dpx}. Assuming (at the lowest order) that the noise is not correlated between different detectors\footnote{See~\cite{Caporali:2025mum} for a study of the impact of correlated seismic noise on the reconstruction of the SGWB with ET.}, this procedure gives access to the signal that would otherwise be buried in, or be indistinguishable from, the detector noise~\cite{LIGOScientific:2019vic}. 

In such searches, the data streams are typically cut into segments, cross-correlated on a segment-by-segment basis, and the result is averaged over segments to output an estimate of the cross-correlation statistic. This procedure greatly simplifies the likelihood function and reduces the computational cost of its evaluation. Indeed, due to the Central Limit Theorem (CLT), when a large number of segments is considered, the estimator for the cross-correlation statistic is expected to follow a Gaussian distribution. Moreover, by segmenting the data stream into short segments, we can more safely assume stationarity to hold within each segment, which allows us to neglect correlations between different frequencies. Despite all these benefits, following this procedure requires some assumptions (e.g., a sufficiently large number of segments is used to construct the estimator to achieve Gaussianity) whose validity and impact must be checked thoroughly. This becomes more important for future detectors, where, in addition to the SGWB, numerous resolvable sources will be present in the data streams, and they will need to be characterized and subtracted.

In this paper, we aim to characterize the biases introduced by the Gaussian likelihood assumption and identify the regimes in which it remains a valid approximation for different GW detectors. We also propose alternative likelihoods that may offer a more accurate description of the underlying statistics. We provide a simple quantitative comparison between these approaches in order to improve the robustness of the SGWB parameter estimation and guide the analysis strategies of future GW detectors. To highlight the results of our work, in~\cref{tab:summary_table}, we summarize all the likelihoods considered in this work, reporting the degree of compression in frequency or number of segments, the assumptions concerning noise characterization, the expected accuracy and precision, and mention some additional features as needed. 

The paper is organized as follows: We begin in~\cref{sec:setting_up} by setting up the full frequency-domain framework, introducing the covariance matrices, data-compression steps, and the exact Whittle Likelihood. Then, in~\cref{sec:one-detector}, we consider a simple ``toy'' example consisting of a single detector and single frequency bin, to illustrate how the Gaussian approximation can break down when the number of segments is small, and to motivate other likelihoods.~\Cref{sec:multi-detectors} generalizes the analysis to multiple detectors (including both white and colored signal and noise models), deriving both the exact and reduced Whittle forms, the Fisher (Gaussian) approximation, and optimal estimators for cross-correlation. In~\cref{sec:realistic_detectors}, we compute statistical uncertainties and systematic biases across segmentation choices for realistic detectors like current LVK, future ET/CE, LISA, and pulsar timing arrays detectors. Moreover, we briefly discuss the situation in anisotropic SGWB searches. Finally, in~\cref{sec:conclusions} we draw our conclusions and outline some future directions for our work. Finally, in~\cref{sec:appendix_tests}, we perform further tests on the likelihoods presented in this work for different signal regimes, number of frequencies, and number of segments.

\begin{table}[t]
\centering
\renewcommand{\arraystretch}{1.4}
\begin{tabular}{|l|c|c|c|c|c|c|c|}
\hline
 Likelihood & $N_s$-compressed & $N_f$-compressed & Noise agnostic  & Unbiased & Coverage & Further info\\
\hline
$\mathcal{L}^{\rm W}(\theta\vert D^s_k)$, ~\cref{e:LW-full}     & \xmarkred & \xmarkred & \xmarkred & \cmarkgreen & \cmarkgreen& - \\
${\cal L}^{\rm G}(\theta\vert D_k)$,~\cref{e:pe_ms:likelihood}     & \cmarkgreen & \xmarkred & \cmarkgreen & \xmarkred & \cmarkgreen& - \\
${\cal L}^{\rm WMAP}(\theta\vert D_k)$,~\cref{e:LWMAP}    & \cmarkgreen & Depends & \xmarkred & \cmarkgreen & \cmarkgreen& See~\cite{Caprini:2019pxz, Flauger:2020qyi},  for $N_f$ compression \\

$\mathcal{L}^{\rm W}(\theta\vert D_k)$,~\cref{e:lnLWishart}     & \cmarkgreen & \xmarkred & \xmarkred &  \cmarkgreen      & \cmarkgreen     & - \\
$\mathcal{L}^{\rm F}(\theta\vert D_k)$,~\cref{e:lnLFisher}     & \cmarkgreen & \cmarkgreen     & \xmarkred & \cmarkgreen     & \cmarkgreen    & Gaussian uncertainties in $\theta$ \\
$\mathcal{L}^{\rm W,\, red}(\theta\vert \bar{D}_k)$,~\cref{e:lnLReduced}  & \cmarkgreen & \xmarkred & \cmarkgreen &  Depends   & \xmarkred (under)  & bias depends on $\bar P_{II,k}$ estimates \\
$\mathcal{L}^{\rm G,\, cc}(\theta\vert \bar{D}_k)$,~\cref{e:lnLGaussian}    & \cmarkgreen & \xmarkred & \cmarkgreen & Depends  & \xmarkred (under)    & bias depends on $\bar P_{II,k}$ estimates \\
$\mathcal{L}^{\rm G, opt}(\theta\vert \bar{D}_k)$,~\cref{e:LGcc_colored}    & \cmarkgreen & \cmarkgreen & \cmarkgreen & Depends & \xmarkred (under)   & bias depends on $\bar P_{II,k}$ estimates \\
\hline
\end{tabular}
\caption{This table summarize all the likelihoods considered in this work, highlighting the assumptions, the level of compression in frequency $f$ or number of segments $\Ns$, the noise assumptions, and the estimation accuracy, together with some implications. For each likelihood, we report the corresponding equations in the main text. We use the symbols ``\cmarkgreen, \xmarkred'' to indicate  the property of the likelihoods, which sometime can depend on some features specified in the most right column. For more details, see main text.  \label{tab:summary_table}}
\end{table}

\section{Setting up the framework}
\label{sec:setting_up}

Schematically, we can model the GW detector output in the frequency domain by writing the data stream in the channel (or detector) $I=1,2,\ldots, \Nd$ as $\tilde{d}_{I,k} =  \tilde s_{I,k} + \tilde n_{I,k}$, where $k$ denotes the discrete frequency bin $f_k\equiv k/T$, with $k=0,1,\ldots, N/2-1$ (non-negative frequencies) and $T\equiv N\Delta t$ is the duration of the observation. 
Here, $\tilde s_{I,k}$ represents the GW signal, while $\tilde n_{I,k}$ refers to the detector/channel noise component. 
A common assumption in GW data analysis is that the signal and noise components are zero-mean Gaussian processes, uncorrelated with each other: i.e., $\langle \tilde s \rangle = \langle \tilde n \rangle = \langle \tilde s \tilde n \rangle = 0$. 
In addition, the noise is often taken to be stationary and uncorrelated across distinct frequency bins. 

Under these Gaussian and stationarity assumptions, the joint probability distribution of the data $\tilde d\equiv \{\tilde{d}_{I,k}\}$ can be written explicitly. In particular, one can write the log-likelihood function for a set of frequency-domain measurements as%
\footnote{In this paper, we will denote the likelihood function as $\mathcal{L}({\rm params}|{\rm data})$, interpreting the model parameters as the variable quantities and the data as fixed (corresponding to a particular measurement).  Alternatively, we could have used the notation $p({\rm data}|{\rm params})$ for the likelihood function, interpreting the data as random variables, conditioned on a fixed set of parameter values.
This latter form of the likelihood function is important if one needs to simulate data to test a data analysis pipeline, for example. 
Since we will be primarily interested in maximum-likelihood estimates and posterior distributions for the parameters, we will ignore additive terms to the log-likelihood function that are independent of the parameter values, even if these terms depend on the data.
But beware that these data-dependent terms are important for evidence and Bayes factor calculations if {\it different} likelihood functions are used to describe the different models.  
If the same likelihood function is used for the different models, then the data-dependent terms will cancel out of the Bayes factor calculation.} 
\begin{equation}
\begin{aligned}
-2\ln \mathcal{L}^{\rm W}(\theta\vert \tilde{d})=&\sum_{k=0}^{N/2-1} \bigg \{2\ln \det \left[2\pi C_{k}(\theta)\right ]
+ \sum_{I,J=1}^{N_{\rm det}}
\tilde{d}^*_{I,k} [C_{k}^{-1}(\theta)]_{IJ} \tilde{d}^{\phantom{*}}_{J,k}
\bigg \} \,.
\label{e:likelihood}
\end{aligned}
\end{equation}
Here $C^{-1}_{k}(\theta)$ denotes the inverse of the $\Nd\times \Nd$ covariance matrix $C_{k}(\theta)$ (for a specific discrete frequency bin $f_k$), whose components can be expressed as 
\begin{equation}
C_{IJ,k}(\theta) 
= \frac{T}{4}P_{IJ,k}(\theta)\,,
\label{e:CIJk}
\end{equation}
where $P_{IJ,k}(\theta)$ are the (1-sided) auto- and cross-power spectral densities (PSDs) of the data
\begin{equation}
\langle \tilde d_{I,k}\rangle =0\,,
\qquad
\langle \tilde d_{I,k} \tilde d_{J,k'}^*\rangle =\frac{T}{2}\delta_{kk'}P_{IJ,k}(\theta)\,.
\label{e:dtilde_moments}
\end{equation}
These power spectral densities can be split into signal and noise components ${P_h}$ and ${P_n}$
\begin{equation}
P_{IJ,k}(\theta)\equiv {P_h}_{IJ,k}(\theta_h) + {P_n}_{IJ,k}(\theta_n)\,,
\end{equation}
which depend on some signal and noise parameters, $\theta_h$, $\theta_n$, respectively. 
In the literature,~\cref{e:likelihood} is also known as the Whittle likelihood~\cite{Moran1951HypothesisTI}. 

It is often convenient to write the log-likelihood in a slightly more compact form by defining the quadratic data combination matrices
\footnote{The real part of the product $\tilde d^{\phantom{*}}_{I,k}\tilde d^*_{J,k}$ enters the trace in~\cref{e:lnLW-compressed}, since $P_{IJ,k}$ is a real symmetric matrix on account of the summation in~\cref{e:likelihood} ranging over only non-negative frequencies.}
\begin{equation}
D_{IJ,k} \equiv \frac{2}{T}{\rm Re}[\tilde d^{\phantom{*}}_{I,k}\tilde d_{J,k}^*]\,.
\label{e:Dtilde}
 \end{equation}
In terms of $D_{IJ,k}$ and $P_{IJ,k}$, we have
\begin{equation}
\begin{aligned}
-2 \ln \mathcal{L}^{\rm W}(\theta\vert D)
=2\sum_{k}\big\{
&\ln \det[P_{k}(\theta)] +{\rm Tr}[P^{-1}_{k}(\theta) D^{\phantom{1}}_{k}]\big\} \,, 
\end{aligned}
\label{e:lnLW-compressed}
\end{equation}
where we have ignored any additive terms that are independent of $\theta$, noting that $D_{IJ,k}$ is a sufficient statistic for this likelihood.%
\footnote{Since the data matrices $D_{IJ,k}$ follow a Wishart distribution~\cite{wishart}, the log-likelihood function given in~\cref{e:lnLW-compressed} is also called the Wishart likelihood.}
Equation~\eqref{e:lnLW-compressed} is a useful starting point for data compression, which is possible, e.g., if the data can be combined over frequency or segments (as discussed in some simple examples below).

When analyzing data from current ground-based detectors (i.e., LVK), each data channel corresponds to a different detector. Since the noises for different detectors are expected to be uncorrelated%
\footnote{Noise correlation can affect GW detectors in different ways: In the triangular configuration, colocated detectors share correlated seismic noise—both Rayleigh and body waves—along with local gravitational-field fluctuations (Newtonian noise).
These effects can persist up to tens of Hz, posing a challenge to measuring an isotropic SGWB.
Separated interferometers are affected by broadband environmental noise, originating from Schumann resonances, which leads to coherent magnetic-field fluctuations~\cite{Thrane-et-al:2013, Coughlin-et-al:2018}.
In this paper, we disregard both these correlated noise effects.}, we have
\begin{equation}
    {P_n}_{{IJ},k}(\theta_n) = \delta_{IJ} {P_n}_{I,k}(\theta_n)\,,
\end{equation}
where ${P_n}_{I,k}$ denotes the noise power spectral density of the $I$-th detector evaluated at frequency bin $f_k$. 
On the other hand, the gravitational-wave background is correlated across multiple detectors, so that
\begin{equation}
    {P_h}_{{IJ},k}(\theta_h) = \Gamma_{IJ}(f_k) {S_h}(f_k;\theta_h)\,.
\label{e:Ph}
\end{equation}
Here, $\Gamma_{IJ}(f)$ denotes the overlap reduction function~\cite{Flanagan:1993, Christensen:1997, Romano:2016dpx} (which encodes the reduction in sensitivity when correlating physically separated and misaligned detectors) and 
\begin{equation}
S_h(f) \equiv\frac{3 H_0^2}{2\pi^2} \frac{\Omega_{\rm gw}(f)}{f^3}\,,
\label{e:Sh}
\end{equation}
is the (one-sided) strain power spectral density for the GWB, with $\Omega_{\rm gw}(f)$ denoting the fractional energy-density spectrum in gravitational waves~\cite{Romano:2016dpx}.

When data are acquired over an extended period, the instrumental noise properties can change significantly over time, so the assumption of strict stationarity cannot be used. For this reason, the data stream is typically divided into a number $\Ns$ of much shorter segments, and the likelihood is expressed as a product
\begin{align}
    \ln \mathcal{L}^{\rm W}
    (\theta\vert D)
    =
    \sum_{s=1}^{\Ns}  \ln \mathcal{L}^{\rm W}
    (\theta^s \vert D^{s}) , 
\label{e:likelihood_total}
\end{align}
where $s$ indexes the segments. 
Note that the parameter vector $\theta^s\equiv (\theta_h,\theta_n^s)$ now carries a segment index, since the noise parameters will generally change from segment to segment.  
This increases the total number of noise parameters that would need to be inferred from the data.
Thus, the most general expression for the log-likelihood function involving multiple channels (detectors), time segments, and frequencies is
\begin{equation}
\begin{aligned}
-2 \ln \mathcal{L}^{\rm W}
(\theta\vert D)=2\sum_s\sum_k \big \{
&\ln \det[P_{k}(\theta^s)]
+ {\rm Tr}[P_{k}^{-1}(\theta^s) D^s_{k}]\big \} \,, 
\label{e:LW-full}
\end{aligned}
\end{equation}
where the quadratic data combination given in eq.~\eqref{e:Dtilde} now depends on the data segment, $D^s_{IJ,k}$.

For current ground-based detectors, the signal is typically too weak to overcome the detector noise, a scenario commonly referred to as the {\it weak-signal} (or {\it noise-dominated}) regime.
In this regime, the individual detector noise power spectra ${P_n^s}_{{I,k}}$ are well-approximated by the detector auto-power $P^s_{II,k}$, and may replaced in the likelihood by (fixed) fiducial estimates $\bar P^s_{II,k}$ obtained, e.g., from spline fits of the quadratic data combinations.
Under the assumption of uncorrelated noise, the GW signal appears uniquely in the cross-correlated data ($I\ne J$).  
Hence, searches for GWBs using ground-based detectors focus primarily on correlating data from pairs of distinct detectors~\cite{Allen:1997ad, Romano:2016dpx}.

It is then customary to define for each segment $s$ the following cross-correlation estimator of the fractional energy-density spectrum
\begin{equation}
\hat\Omega^s_{IJ,k}\equiv
\left(\frac{2\pi^2}{3H_0^2}\right)
\frac{f_k^3}{\Gamma_{IJ}(f_k)}D^s_{IJ,k}\,,
\end{equation}
where $I\ne J$.
The numerical factors on the rhs are chosen so that $D^s_{IJ,k}/\Gamma_{IJ}(f_k)$ is an estimator of $S_h(f_k)$ (see eqs.~\eqref{e:dtilde_moments}, \eqref{e:Dtilde}, \eqref{e:Ph}), while $f_k^3(2\pi^2/3H_0^2)$ converts strain power $S_h(f_k)$ to energy-density spectrum units (see~eq.~\eqref{e:Sh}).  
In the weak-signal regime, one can further combine these segment-dependent estimators by weighting them by the inverse of the (approximate) variances
\begin{equation}
(\sigma^2){}^{s}_{IJ,k}\approx 
\left(\frac{2\pi^2}{3H_0^2}
\frac{f_k^3}{\Gamma_{IJ}(f_k)}\right)^2
\bar P^s_{II,k}\bar P^s_{JJ,k}\,.
\end{equation}
If the number of segments is sufficiently large, this averaging over segments yields data combinations $\hat\Omega_{IJ,k}$ that (according to the central limit theorem) are approximately Gaussian distributed.
In this case, 
\be
\begin{aligned}
- 2 \ln {\cal L}^{\rm G} (\theta_h\vert \hat\Omega) =
& \sum_{I<J}\sum_k\left\{
\ln (2\pi \sigma_{IJ,k}^2)
+\frac{\left[\hat\Omega_{IJ,k} - \Omega_{\rm gw}(f_k, \theta_h)\right]^2}{\sigma^2_{IJ,k}}\right\}\,,
\label{e:pe_ms:likelihood}
\end{aligned}
\ee
where $\Omega_{\rm gw}(f, \theta_h)$ is the model spectrum and
\begin{equation}
1/\sigma^2_{IJ,k}\equiv \sum_s 1/(\sigma^2)^s_{IJ,k}
\end{equation}
is the variance of the estimator $\hat\Omega_{IJ,k}$.
Note that the above summation is over only distinct pairs of detectors $I\ne J$, and that the only parameters we are trying to infer from the data are those for the gravitational-wave background $\theta_h$. 

Following this procedure significantly reduces the amount of data, leading to a significant reduction of the computational cost, which, in turn, speeds up the analysis. 
However, with the enhanced sensitivity expected from future GW detectors,  the precision of these approximations might become insufficient due to improved measurement accuracy. In the following sections, we explore the conditions under which this approximate likelihood yields reliable maximum likelihood (ML) estimates, as well as scenarios in which ML biases are likely to become evident. 

\section{One detector case}
\label{sec:one-detector}

We start by presenting a very simple one-detector analysis, treating each frequency bin independently, which requires no assumptions on the frequency dependence of the signal and noise. We postpone the more complex scenario, where several frequency bins are combined, to the following sections.

\subsection{A simple example}\label{sec:toymodel}

As a very simple example, consider $\Ns$ segments of stationary Fourier-domain data $\tilde d^s_{I,k}$ from a single detector $I$, and focus attention on a single frequency bin $f_k$.
We will assume that the segments are statistically equivalent and independent of one another.
Since $I$ and $k$ are fixed for this example, we will drop them as subscripts and write $\tilde d_s$ for the data and $D_s\equiv 2|\tilde d_s|^2/T$ for the quadratic data combination~\cref{e:Dtilde}, where $s=1,2,\ldots, \Ns$.
The frequency domain data satisfy the quadratic expectation values
\begin{equation}
\langle \tilde d_s\rangle =0\,,
\qquad
\langle \tilde d_s \tilde d_{s'}^*\rangle =
\frac{T}{2}\delta_{ss'}\,P\,,
\label{e:dtilde_moments_ex1}
\end{equation}
where our assumption regarding stationarity across segments means that the value $P$ of the power spectrum at frequency $f_k$ is independent of the segment $s$.
For this example, we will take $P$ as the parameter we want to infer from the data. Note that, at this level, we do not try to separate signal and noise, meaning $P$ represents the total autocorrelated power.

The normalized probability distribution for the $\Ns$ data samples $\tilde d\equiv\{\tilde d_s|s=1,2,\ldots,\Ns\}$ is then
\begin{equation}
    p(\tilde{d}|P) \equiv \left(\frac{4}{2\pi T P}\right)^{\Ns}\exp\left[-\frac{1}{2}\sum_s\frac{4|\tilde{d}_s|^2}{TP}\right]\,.
\label{e:p(dtilde|P)}
\end{equation}
Note that the argument of the exponential can be written more simply in terms of the quadratic data combination 
\begin{equation}
D\equiv \frac{1}{\Ns} \sum_s D_s \equiv \frac{1}{\Ns}\sum_{s}\frac{2}{T}|\tilde{d}_s|^2\,,
\label{e:D-1d}
\end{equation}
where we compressed $D_s$ over segments.
Indeed, one can show that $D$ is the ML estimator of $P$, which we will denote as 
\begin{equation}
\ML{P}\equiv D\,,
\label{e:PML}
\end{equation}
and has mean and variance
\begin{equation}
\langle D\rangle = P\,,\qquad
\langle D^2\rangle-\langle D\rangle^2 = \frac{P^2}{\Ns}\,.
\label{e:mean_var_Phat}
\end{equation}
The latter can be calculated directly from~\cref{e:dtilde_moments_ex1} using Isserlis's (or Wick's) theorem~\cite{isserlis:1918} to expand the expectation value of a product of four $\tilde d_s$'s.
$D$ is obviously non-Gaussian, being proportional to a sum of squared Gaussian random variables. In particular, it follows a scaled chi-squared distribution, with scaling factor $P/2\Ns$ and $2\Ns$ real degrees of freedom. 

\begin{figure}[t!]
\centering
\includegraphics[width=0.6\textwidth]{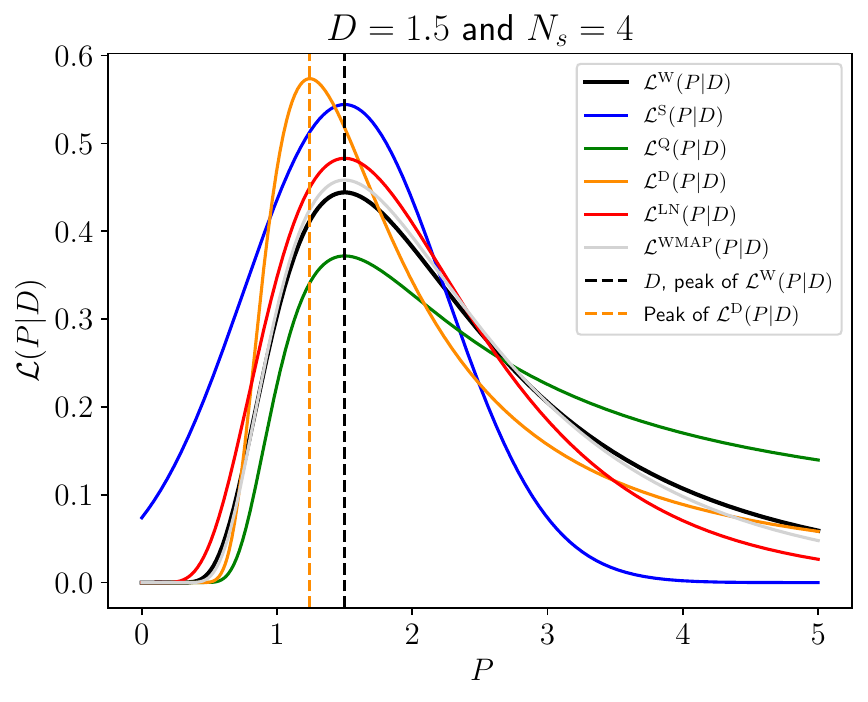}
\centering
\caption{Plots of ${\cal L}^{\rm W}(P\vert D)$ (black) together with several approximate likelihood functions (all normalized to have unit area), plotted as a function of the unknown parameter $P$ for fixed data $D$.
For this particular example, $D=1.5$ and the number of data segments is $\Ns=16$.}
\label{fig:lnLike}
\end{figure}

By inspection of eqs.~\eqref{e:p(dtilde|P)} and \eqref{e:D-1d}, one sees that the quadratic data combination $D$ is a sufficient statistic for~\cref{e:p(dtilde|P)}.
Starting from eq.~\eqref{e:p(dtilde|P)}, and making a change of variables from the original dataset $\{\tilde d_s|s=1,\ldots,\Ns\}$ to $D$, one obtains the properly normalized conditional probability distribution for $D$ given $P$ as 
\begin{equation}
p^{\rm W}(D|P) = \frac{\Ns^{\Ns} D^{\Ns-1}}{\Gamma(\Ns)}\frac{e^{-\Ns D /P}}{P^{\Ns}}\,,
\label{e:p(Phat|P)}
\end{equation}
where $\Gamma(x)$ is the Gamma function.
The corresponding log-likelihood function (ignoring additive terms that are independent of the parameter $P$) is simply: 
\begin{equation}
-2\ln{\cal L}^{\rm W}(P|D)
=2\Ns(\ln P + D/P)\,.
\label{e:LW}
\end{equation}
A plot of ${\cal L}^{\rm W}(P\vert D)$ is given in~\cref{fig:lnLike} (black curve) together with several other approximate likelihood functions, which we discuss in the next two subsections.

\subsection{Approximate Gaussian likelihood functions}

Using the mean and variance of $D$ given in eq.~\eqref{e:mean_var_Phat}, one can construct different {\it approximate Gaussian-distributed} likelihood functions, depending on how we want to treat the variance term.
Following the notation in~\cite{Hamimeche-Lewis:2008}, we consider the following ``Gaussian-like'' approximations to the Whittle likelihood function.
Plots of these likelihood functions are given in~\cref{fig:lnLike}, all normalized to have unit area.
Thus, one can think of these plots as {\it posterior probability distributions} for $P$, calculated assuming a flat prior probability distribution.

(i) The {\it symmetric} Gaussian approximate likelihood function is defined by
\begin{equation}
-2\ln{\cal L}^{\rm S}(P|D)
=\Ns\left(\frac{D-P}{D}\right)^2\,,
\label{e:LS}
\end{equation}
where the theoretical variance $P^2/\Ns$ is replaced by the data-dependent expression $D^2/\Ns$.
This likelihood peaks at $D$ (which is the ML estimate of $P$ from the full Whittle likelihood) and is symmetric about around the peak.

(ii) Alternatively, one can define the {\it quadratic} Gaussian approximate likelihood function, whose logarithm is quadratic in the data combination $D$:
\begin{equation}
-2\ln{\cal L}^{\rm Q}(P|D)
=\Ns\left(\frac{D-P}{P}\right)^2\,.
\label{e:LQ}
\end{equation}
This approximate likelihood uses the theoretical expression for the variance but doesn't include the standard log-determinant contribution from the variance (which is used below).
This likelihood again peaks at $D$, but it is no longer symmetric around the peak.

(iii) Finally, we can consider the standard Gaussian likelihood function, which uses theoretical expressions for the variance and includes the log-determinant term (hence the `D' superscript for {\it determinant}):
\begin{equation}
-2\ln{\cal L}^{\rm D}(P|D)
=\Ns\left(\frac{D-P}{P}\right)^2 + 2\ln P \,.
\label{e:LD}
\end{equation}
The shape of this likelihood agrees more closely with the Whittle likelihood than those associated with the other two Gaussian approximations, but its peak does not agree with $D$.
By maximizing ${\cal L}^{\rm D}(P|D)$ with respect to $P$, one can show that ${\cal L}^{\rm D}$ is peaked at 
\begin{equation}
\begin{aligned}
\hat{P}^{\rm D}&\equiv
D\left(\sqrt{1+{4}/{\Ns}}-1\right)\Ns/2
= D\left[1-{1}/{\Ns} + {\cal O}\left({1}/{\Ns^2}\right)\right]\,. 
\end{aligned}
\end{equation}
This difference is indicated in~\cref{fig:lnLike} by the dashed orange vertical line.
Since $\langle D\rangle=P$, it follows that
\begin{equation}
\langle\hat P^{\rm D}\rangle = P\left[1-{1}/{\Ns} + {\cal O}\left({1}/{\Ns^2}\right)\right]\,,
\end{equation}
which means that $\hat P^{\rm D}$ (the ML estimator of $P$ associated with the Gaussian likelihood $\cal L^{\rm D}(P|D)$) is a {\it biased} estimate of the parameter $P$.
For this case, the fractional bias factor is given by
\begin{equation}
\mathcal{B}\equiv
\frac{\langle\hat{P}^{\rm D}\rangle - P}{P}\simeq - \frac{1}{\Ns}\,,
\label{e:bias}
\end{equation}
which shows that the Gaussian-based estimator systematically underestimates the true variance by a factor $1/\Ns$. This bias will clearly diminish in the limit where $\Ns\gg 1$. 

We further illustrate this with a slightly more complicated example. 
Consider data dominated by a single 
SGWB component parametrized by $P$, measured over $N_f$ frequency bins and averaged across $N_s$ segments. 
Under the Gaussian approximation, one expects that by sampling over a total number of $N_f\times N_s$ independent realizations, the fractional statistical uncertainty of the ML estimator $\ML{P}$ of the full Whittle likelihood around its true value $P$ will scale as
\begin{equation}
{\cal S}\equiv\Delta \ML{P} 
\equiv \left|\frac{\ML{P}-P}{P}\right|
\simeq \frac{1}{\sqrt{\Nf \Ns}}\,.
\end{equation}
The number of frequencies can be roughly estimated from the duration of the observational run $T$ and the maximum frequency  $f_{\rm max}$ as $N_f \sim f_{\rm max} T$. Note that, depending on the values of  $N_f$ and $N_s$, the resulting statistical uncertainty ${\cal S}$ may be smaller than the intrinsic bias $\mathcal{B}$. In such a case, the bias can become relevant and affect the reconstruction. Furthermore, in realistic scenarios where detector noise is present and we are not in a signal-dominated regime, a practical criterion to estimate ${\cal S}$ would be to count only frequency bins where the noise PSD is much smaller than the signal power, in order to improve the estimator. In the opposite case, the presence of the noise would increase the level of uncertainty.

\subsection{A CMB-inspired alternative likelihood}

The Gaussian-approximate likelihood functions described in the previous subsection are quadratic in the fractional difference 
\begin{equation}
\delta\equiv \frac{D-P}{P}
\quad\Leftrightarrow\quad D=P(1+\delta)\,.
\end{equation}
It turns out that by considering a linear combination of the quadratic Gaussian likelihood $\ln{\cal L}^{\rm Q}(P\vert D)$ defined in eq.~\eqref{e:LQ}, and a {\it log-normal} likelihood
\begin{equation}
-2 \ln {\cal L}^{\rm LN}(P\vert D)=
\Ns\left[\ln\left(\frac{D}{P}\right)\right]^2\,,
\label{e:LLN}
\end{equation}
we can construct an approximate likelihood that agrees with the full Whittle likelihood ${\cal L}^{\rm W}(P\vert D)$ given in eq.~\eqref{e:LW} to third-order in $\delta$.

To demonstrate this, we follow the discussion presented in~\cite{WMAP:2003pyh} in the context of CMB physics.
We start by rewriting the Whittle likelihood ${\cal L}^{\rm W}(P\vert D)$ by adding to the rhs of~\cref{e:LW} a couple of terms that are independent of the parameter $P$ and hence do not change the shape of ${\cal L}^{\rm W}(P\vert D)$ viewed as a function of $P$:
\begin{equation}
-2\ln{\cal L}^{\rm W}(P|D)
\simeq 2\Ns\left[\ln \left(\frac{P}{D}\right) + \frac{D}{P}-1\right]\,.
\end{equation}
In terms of $\delta$, we have
\begin{equation}
\begin{aligned}
-2 \ln {\cal L}^{\rm W}(P\vert D)
&=2\Ns\left[-\ln(1+\delta) + (1+\delta)-1\right]
=\Ns\left[\delta^2-\frac{2}{3}\delta^3+{\cal O}(\delta^4)\right]\,.
\end{aligned}
\end{equation}
Similarly, we can express ${\cal L}^{\rm Q}$ and ${\cal L}^{\rm LN}$ in terms of $\delta$:
\begin{equation}
-2 \ln {\cal L}^{\rm Q}(P\vert D)
=\Ns\,\delta^2
\end{equation}
and
\begin{equation}
\begin{aligned}
-2 \ln {\cal L}^{\rm LN}(P\vert D)
&=\Ns\left[\ln(1+\delta)\right]^2
= \Ns\left[\delta^2-\delta^3+{\cal O}(\delta^4)\right]\,.
\end{aligned}
\end{equation}
In this way, the full Whittle likelihood can be written as
\begin{equation}
\ln {\cal L}^{\rm W}(P\vert D)
=\ln {\cal L}^{\rm WMAP}(P\vert D)
+{\cal O} (\delta^4),
\label{e:likelihoodform}
\end{equation}
where the WMAP likelihood approximation is defined as
\begin{equation}
\ln {\cal L}^{\rm WMAP}(P\vert D)
\equiv
\frac{1}{3} \ln {\cal L}^{\rm Q}(P\vert D)
+\frac{2}{3}\ln {\cal L}^{\rm LN}(P\vert D)\,.
\label{e:LWMAP}
\end{equation}
As shown in~\cref{sec:toymodel}, the fractional deviation $\delta $ has a typical magnitude that scales as ${\cal O}(\Ns^{-1})$  once the data are averaged over $\Ns$ (statistically independent) segments. The next correction---arising from the quartic term in the expansion---is ${\cal O}(\Ns^{-2})$ and therefore subdominant for practical GWB searches.
As evident from inspection of~\cref{fig:lnLike} (compare grey and black curves), the WMAP likelihood approximation agrees very closely with that of the full Whittle likelihood function. 

\section{More detectors}
\label{sec:multi-detectors}

We now consider a series of more complicated examples, which involve multiple detectors having either white or colored signal and noise models.
For all cases, we will assume that the data are taken over a total observation period $T_{\rm obs}$, and are split into $\Ns$ segments of duration $T\equiv T_{\rm obs}/\Ns$.
We will assume that the data are stationary across segments, which means that the parameters $\theta$ describing the signal and noise are not segment dependent.
This assumption will allow us to combine the data over segments as was done in~\cref{sec:one-detector}.
We will also assume that individual frequency bins are statistically independent, which is a good approximation when the data are correlated over time scales much shorter than the segment duration $T$.

If we additionally assume that the signal and noise are white with variances $\sigma_h^2$, $\sigma_{n_I}^2$, for $I=1,2,\ldots,\Nd$, then the $C_k(\theta)$ matrices are also independent of frequency since the one-sided power spectral densities are given by
\begin{equation}
\begin{aligned}
P_{n_{I,k}} = 2\sigma^2_{n_I} \,\Delta t\,,\quad 
S_h(f_k) = 2\sigma^2_h\, \Delta t \,,
\label{e:PS-white}
\end{aligned}
\end{equation}
where $\Delta t \equiv T/N$, with $N$ being the number of time domain samples in a single segment.
(Recall that the number of non-negative frequency bins is $\Nf=N/2$.)
Also, assuming that the overlap function $\Gamma_{IJ}(f)$ is independent of frequency and equals unity when $I=J$,
then 
\begin{equation}
C_{k}(\theta) = \frac{T}{4}\,P\,,
\end{equation} 
where the components $P_{IJ}$ of the (constant) $P$ matrix are given by
\begin{equation}
P_{IJ}= \left[
\begin{matrix}
P_1 & \Gamma_{12} S_h & \cdots & \Gamma_{1 \Nd} S_h\\
\Gamma_{12} S_h & P_2 &\cdots & \Gamma_{2\Nd} S_h\\
\vdots & \vdots & \ddots & \vdots \\
\Gamma_{1 \Nd} S_h & \Gamma_{2\Nd} S_h & \cdots & P_{\Nd}
\end{matrix}
\right]\,,
\label{e:PIJ}
\end{equation}
with
\begin{equation}
P_I\equiv P_{II} \equiv P_{n_I} + \Gamma_{II}\,S_h=2\,(\sigma^2_{n_I} + \sigma^2_h)\Delta t\,.
\label{e:PI-white}
\end{equation}
For this white signal and noise case, we will take $\theta\equiv (P_1,\ldots, P_\Nd, S_h)$ as the set of parameters that we want to infer from the data; they are simply related to $\sigma^2_h$ and $\sigma^2_{n_I}$ using equations~\eqref{e:PS-white} and~\eqref{e:PI-white}.

Making the above simplifying substitutions into eq.~\eqref{e:LW-full}, we obtain
\begin{equation}
\begin{aligned}
-2 \ln \mathcal{L}^{\rm W}
(\theta\vert D)=2\Ns\Nf\left(\ln \det(P)
+ {\rm Tr}[P^{-1}D]\right) \,, 
\label{e:LWC}
\end{aligned}
\end{equation}
where the quadratic data matrix $D$ is 
\begin{equation}
D_{IJ} \equiv \frac{1}{\Ns\Nf}\sum_s\sum_k D_{IJ,k}^s\,,
\label{e:DIJ}
\end{equation}
and where we have ignored terms on the rhs of~\cref{e:LWC} that are independent of the parameters $\theta$.
Note that~\cref{e:LWC} is a {\it compressed}-version of~\cref{e:LW-full}, since the data combination $D_{IJ}$ above has been compressed over both segments and frequency bins given the assumptions of stationarity across segments and white signal and noise models.
These equations will be the starting point for the analyses in~\cref{sec:2d} and~\cref{sec:3d}. All frequency-dependent considerations are postponed to~\cref{sec:nonwhite}.

\subsection{2-detectors, white signal, white noise}
\label{sec:2d}

For the case of 2-detectors with white signal and white noise, the log-likelihood function given in eq.~\eqref{e:LWC} takes the following explicit form:
\begin{equation}
\begin{aligned}
&-2\ln{\cal L}^{\rm W}(\theta|D) = 2\Ns\Nf\left[\ln(P_1P_2-\Gamma_{12}^2 S_h^2) +\left(1-\frac{\Gamma_{12}^2 S_h^2}{P_1P_2}\right)^{-1}
\left(\frac{D_{11}}{P_1} + \frac{D_{22}}{P_2} - 2 \frac{D_{12}\Gamma_{12} S_h}{P_1 P_2}\right)\right]\,.
\label{e:LWC-2d}
\end{aligned}
\end{equation}
If we maximize this log-likelihood function with respect to the parameters $\theta\equiv (P_1, P_2, S_h)$, we obtain the following ML estimates
\begin{equation}
\hat P_1 = D_{11}\,,\quad
\hat P_2 = D_{22}\,,\quad
\hat S_h = \frac{D_{12}}{\Gamma_{12}}\,.
\label{e:ML-2d}
\end{equation}
These estimators are quadratic in the Fourier domain data $\tilde d_{I,k}$.
In general, ML estimators of the parameters need not be quadratic, as we shall see for the 3-detector examples studied in~\cref{sec:3d} and~\cref{sec:nonwhite}.

The variance of these estimators can be computed starting from the quadratic expectation values of the Fourier-domain data
\begin{equation}
\langle\tilde d_{I,k}^s \tilde d_{J,k'}^{s'*}\rangle = \frac{T}{2}\delta_{ss'}\delta_{kk'} P_{IJ}\,.
\end{equation}
Higher-order moments required for the variance calculation can then be evaluated by applying Isserlis's theorem~\cite{isserlis:1918}, which allows one to express fourth-order expectation values of $\tilde d$ in terms of products of second-order moments.
Alternatively, one can obtain the variances from the inverse of the {\it Fisher Information Matrix} (FIM) defined as
\begin{equation}
F_{ij}
\equiv\bigg\langle - \frac{\partial^2 \ln {\cal L}^{\rm W}(\theta\vert D)}{\partial\theta_i\partial\theta_j}
\bigg\rangle\,,
\label{e:F}
\end{equation}
which, for this particular example, takes the form
\begin{equation}
F_{ij} = \Ns\Nf\,{\rm Tr}\,\left[P^{-1} \frac{\partial P}{\partial \theta_i}P^{-1} \frac{\partial P}{\partial \theta_j}\right]\,.
\label{e:F-example}
\end{equation}
The variance for parameter $\theta_i$ is given by
\begin{equation}
\sigma^2_{\theta_i} = (F^{-1})_{ii}\,.
\end{equation}
More generally, the covariance between pairs of parameters $\theta_i$ and $\theta_j$ is given by
${\rm Cov}(\theta_i,\theta_j) = (F^{-1})_{ij}$.
For this example,
\begin{equation}
(F^{-1})_{ij}= \frac{1}{\Ns\Nf}\left[
\begin{matrix}
P_1^2 & \Gamma_{12}^2 S^2_h & P_1 S_h\\
\Gamma_{12}^2 S^2_h & P_2^2 & P_2 S_h\\
P_1 S_h & P_2 S_h & \frac{1}{2}\left(S_h^2 + \frac{P_1 P_2}{\Gamma_{12}^2}\right)
\end{matrix}
\right]\,.
\label{e:Finv-example}
\end{equation}
Note that the variance of $\hat S_h$ is valid here even if the signal power is large compared to that of the noise.

By definition, the FIM and its inverse involve the model parameters, whose values are unknown.
In practice, one evaluates the components of these matrices, $F_{ij}$ and $(F^{-1})_{ij}$, at the ML estimates%
\footnote{ For more complicated examples where the ML estimators of the full Whittle likelihood are not given by simple quadratic combinations of the data, like in~\cref{e:ML-2d}, one can calculate the ML estimators using the iterative FIM method described in~\cref{sec:3d}, see~\cref{e:iterate}.}
given in~\cref{e:ML-2d}.
We will denote these data-dependent expressions with ``hats'', i.e., $\hat{F}_{ij} \equiv F_{ij}|_{\theta=\hat\theta}$ and similarly for $(\hat{F}^{-1})_{ij}$.

Using the ML estimates of the parameters and the associated covariance matrix $\hat F^{-1}$, we can construct a multivariate-Gaussian approximate to the Whittle likelihood function~\cref{e:LWC}, which we will denote as ${\cal L}^{\rm F}(\theta|D)$, with `F' for Fisher.
Ignoring terms that are independent of the parameters $\theta_i$, we define
\begin{equation}
-2\ln {\cal L}^{\rm F}(\theta|D) =\sum_{i,j} (\hat\theta_i-\theta_i)\hat F_{ij}(\hat\theta_j - \theta_j)\,,
\label{e:LF}
\end{equation}
where the indices $i,j$ run over the total number of parameters, and the estimators $\ML{\theta}_i$ are expressed in terms of the data combinations $D_{IJ}$ by~\cref{e:ML-2d}\footnote{See previous footnote regarding calculating $\hat\theta_i$ for the more general scenario.}.
This is a symmetric Gaussian approximation for each parameter similar in spirit to~\cref{e:LS}.
By construction, the ML estimators of the parameters and the magnitude of their uncertainties for this approximate likelihood function agree with those from the full Whittle likelihood~\cref{e:LWC-2d}.

It is also possible to construct a ``reduced'' non-Gaussian likelihood starting with $\ln{\cal L}^{\rm W}(\theta|D)$ in~\cref{e:LWC-2d}, focusing attention on just the GWB parameter $S_h$.
For this case, we replace the parameters $P_1$, $P_2$ and data combinations $2D_{11}/T$, $2D_{22}/T$ by fixed quantities $\bar{P}_1$, $\bar{P}_2$, which are not inferred from the analysis.
Rather, they can be either {\it fiducial estimates} of the auto-power 
constructed, e.g., using spline fits of the detector output, or they may be estimated from the data segment itself.%
\footnote{We do not recommend using data-dependent estimators of the auto-power that are constructed from the analysis segment, since they may lead to biases in parameter estimates, as shown for example in~\cref{fig:3channels_constraints_colored}.
Using power spectrum estimates from neighboring segments avoids this problem, see, e.g.,~\cite{Matas-Romano:2021}.}
These replacements give rise to {\it reduced} power spectrum and quadratic data combination matrices, $\bar P_{IJ}$ and $\bar D_{IJ}$, which differ from $P_{IJ}$ and $D_{IJ}$ defined in~\cref{e:PIJ} and~\cref{e:DIJ} only along the diagonal. 
With these replacements, we can define the {\it reduced Whittle} likelihood function for an arbitrary number of detectors, assuming here that the signal and noise are both white
\begin{equation}
\begin{aligned}
&-2 \ln \mathcal{L}^{\rm W,\, red}
(S_h\vert \bar D;\bar P_1,\cdots,\bar P_{\Nd})
=2\Ns\Nf\left[\ln \det(\bar P)
+ {\rm Tr}(\bar P^{-1}\bar D)\right] \,.
\label{e:LWR}
\end{aligned}
\end{equation}
For two detectors, the above likelihood depends only on the cross-correlated data $D_{12}$ and is conditioned on $\bar{P}_1$ and $\bar{P}_2$.
Explicitly,
\begin{equation}
\begin{aligned}
&-2\ln{\cal L}^{\rm W,\, red}(S_h|D_{12};\bar{P}_1,\bar{P}_2) = 2\Ns\Nf\left[\ln(\bar P_1\bar P_2-\Gamma_{12}^2 S_h^2) +2\left(1-\frac{\Gamma_{12}^2 S_h^2}{\bar{P}_1\bar{P}_2}\right)^{-1}\left(1 - \frac{D_{12}\Gamma_{12} S_h}{\bar{P}_1 \bar{P}_2}\right)\right]\,.
\label{e:Lred-2d}
\end{aligned}
\end{equation}
One can show that this reduced likelihood function is peaked at $D_{12}/\Gamma_{12}$, which is the ML estimator of $S_h$ of the Whittle likelihood function as indicated in~\cref{e:ML-2d}.
Since this likelihood function is conditioned on $\bar{P}_1$ and $\bar{P}_2$, the uncertainty in the recovery of $S_h$ will be underestimated compared to that obtained by marginalizing over $P_1$ and $P_2$.

Alternatively, a reduced likelihood function, which is Gaussian in the cross-correlated data $D_{12}$, can be constructed by again using fiducial estimates $\bar P_1$, $\bar P_2$ for the parameters $P_1$, $P_2$.
One defines (again up to additive terms independent of $S_h$)
\begin{equation}
-2\ln {\cal L}^{\rm G,\,cc}(S_h|D_{12};\bar{P}_1,\bar{P}_2) =\frac{(\hat S_{h,\,{\rm cc}} - S_h)^2}{\sigma^2_{\hat S_{h,\,{\rm cc}}}}\,,
\label{e:LGcc}
\end{equation}
where 
\begin{equation}
\begin{aligned}
\hat S_{h,\, {\rm cc}}\equiv
\frac{D_{12}}{\Gamma_{12}}\,,
\qquad
\sigma^2_{\hat S_{h,\, {\rm cc}}}
\equiv\frac{1}{2\Ns\Nf}\frac{\bar{P}_1 \bar{P}_2}{\Gamma_{12}^2}\,.
\end{aligned}
\end{equation}
This is a symmetric Gaussian likelihood function, quadratic in the data $\tilde d$, depending only on the cross-correlations $D_{12}$.
The corresponding variance is valid only for sufficiently weak GWB signals.
This is evident from inspecting the 33-component of $F^{-1}$ in~\cref{e:Finv-example}, which shows that the signal-signal contribution $S_h^2$ is missing from $\sigma_{\hat S_{h,\,{\rm cc}}}^2$.

\begin{figure}[t!]
\centering
\includegraphics[width=0.6\textwidth]{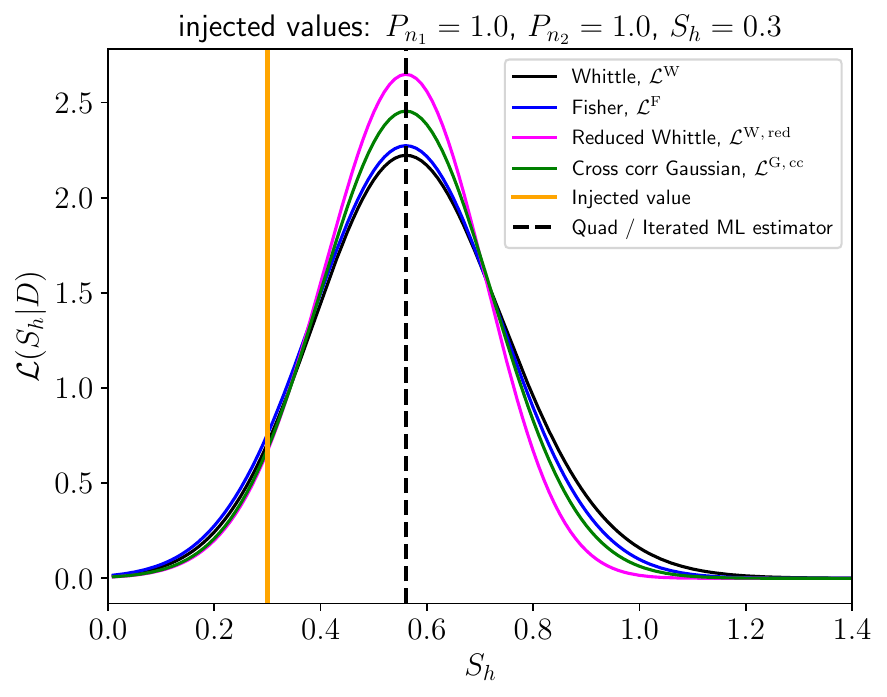}
\centering
\caption{Comparison of the posterior distributions for $S_h$ using the Whittle, Fisher, reduced Whittle, and cross-correlation approximate likelihood functions for the two-detector, white signal, white noise example.
The plots for ${\cal L}^{\rm W}$ and ${\cal L}^{\rm F}$ were marginalized over the parameters $P_1$ and $P_2$.
The marginalization was done via simple numerical integration over a grid of $P_1$ and $P_2$ values, assuming flat priors for all the parameters.
For this analysis, $\Ns=4$ and $\Nf=32$ corresponding to a total compression factor of $\Ns\Nf=128$.}
\label{fig:compare-2d}
\end{figure}

A comparison of the likelihood functions ${\cal L}^{\rm W}$,
${\cal L}^{\rm F}$,
${\cal L}^{\rm W,\, red}$,
${\cal L}^{\rm G,\, cc}$,
plotted as functions of the GW power spectrum parameter $S_h$, is given in~\cref{fig:compare-2d}.
The likelihood functions ${\cal L}^{\rm W}$ and ${\cal L}^{\rm F}$ have been marginalized over the other parameters $P_1$, $P_2$, while ${\cal L}^{\rm W,\, red}$,
${\cal L}^{\rm G,\, cc}$ are conditioned on  fiducial estimates $\bar P_1$, $\bar P_2$ of these parameters.
All the plots have been normalized to have unit area, so they can be interpreted as posterior distributions for $S_h$, under the assumption of flat priors.
The marginalization over $P_1$, $P_2$ for ${\cal L}^{\rm W}$ and ${\cal L}^{\rm F}$ have been done via simple numerical integration as opposed to MCMC methods, which will be used in~\cref{sec:3d} for the 3-detector examples.
For these plots, the simulated data $\tilde d_{I,k}^s$ had $\Ns=4$ and $\Nf=32$ corresponding to a total compression factor of $\Ns\Nf=128$.

\subsection{3-detectors, white signal, white noise}
\label{sec:3d}

We now extend the analyses of the previous subsection to the case of 3-detectors, again assuming stationary, white signal, and white noise power spectra.
The relevant parameters for this problem are the constant power spectral densities $\theta\equiv (P_1, P_2, P_3, S_h)$.
Since the following analysis extends easily to an arbitrary number of detectors having this simple signal+noise model, we will write down the general form of these expressions whenever possible.

\begin{figure}[t!]
\centering
\includegraphics[width=0.6\textwidth]{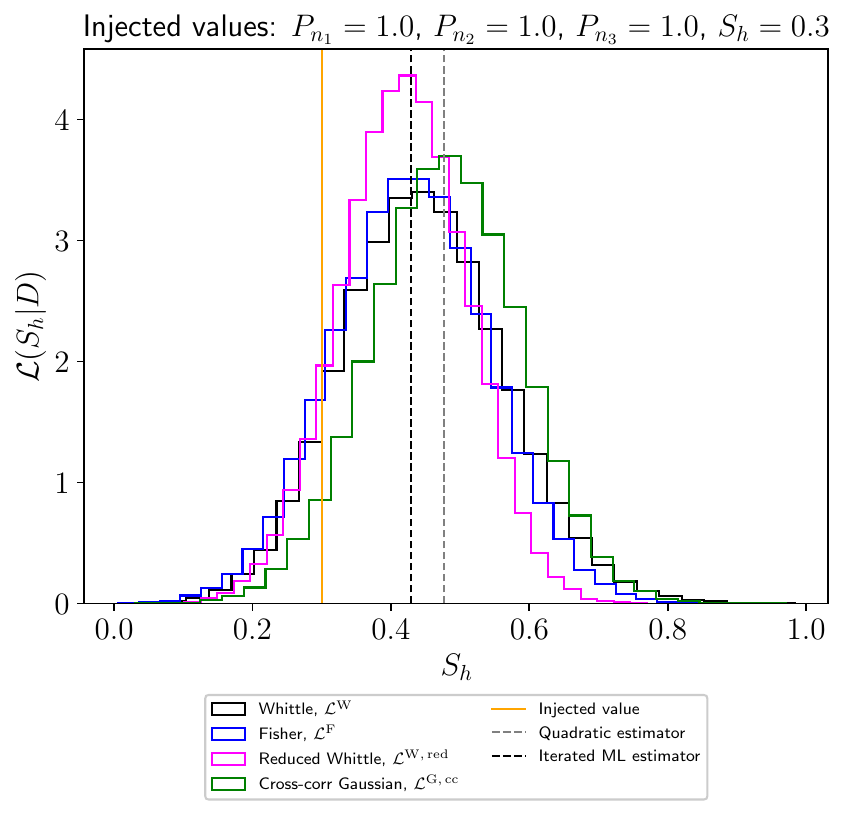}
\centering
\caption{Similar to~\cref{fig:compare-2d}, but for the 3-detector, stationary, white signal, white noise example. 
For this 3-detector example, the marginalizations are done using MCMC methods, assuming flat priors for all the parameters $P_1$, $P_2$, $P_3$, $S_h$.
Note that for 3-detectors, the ML estimate of the parameter $S_h$ does not agree with the quadratic estimator for $S_h$.}
\label{fig:compare-3d}
\end{figure}

The general Whittle likelihood function ${\cal L}^{\rm W}(\theta\vert D)$ for the multi-detector, stationary, white signal, white noise case was already written down in~\cref{e:LWC}.
Similarly, the multivariate-Gaussian likelihood function ${\cal L}^{\rm F}(\theta|D)$ for this case has the same form as that given in~\cref{e:LF}.
However, for three or more detectors, the ML estimators $\hat\theta_i$, $i=1,2, \cdots, \Nd+1$ of the Whittle likelihood in~\cref{e:LWC} are no longer guaranteed to be given by simple extensions of the quadratic estimators~\cref{e:ML-2d}.
Instead, the ML estimators of the parameters $\theta=(P_1,\ldots,P_{\Nd}, S_h)$ can be obtained by iterating the equations
\begin{equation}
\begin{aligned}
&\Delta\ML{\theta}_i^{(n)} = \sum_{j}[F^{-1}(\theta)]_{ij}\frac{\partial \ln{\cal L}^{\rm W}(\theta\vert D)}{\partial\theta_j}\Bigg|_{\theta=\ML{\theta}^{(n)}}\,, 
\\
&\ML{\theta}_i^{(n+1)}=\ML{\theta}_i^{(n)}
+\Delta\ML{\theta}_i^{(n)}\,,
\label{e:iterate}
\end{aligned}
\end{equation}
for $n=0,1, \ldots$ until the absolute value of the deviations between successive iterations $|\Delta\ML{\theta}_i^{(n)}|$ drop below some prescribed tolerance.
The definition of the FIM $F_{ij}$ and its expression for the case of stationary, white signal, and white noise are given in equations~\cref{e:F} and~\cref{e:F-example}.
An explicit expression for the partial derivative of the log-likelihood function entering~\cref{e:iterate} for this simple signal+noise model is 
\begin{equation}
\frac{\partial \ln{\cal L}^{\rm W}(\theta\vert D)}{\partial\theta_i}=\Ns\Nf \, {\rm Tr}\left[P^{-1}\frac{\partial P}{\partial\theta_i}P^{-1}\left(D-P\right)\right]\,.
\end{equation}
Typically, one starts the iteration by assuming that the parameter values are given by the quadratic estimators%
\footnote{Equation~\eqref{e:Sh_quad} for $\hat S_h^{(0)}$ is a simple cross-correlation estimator that is most appropriate when the total auto-correlated power for all detectors is the same.  
If that is not the case, one can construct a smaller variance estimator by forming a $1/\sigma^2$ weighting of $D_{IJ}/\Gamma_{IJ}$ with their variances (which we would either need to know a~priori or approximate using fiducial estimates).
A discussion of optimal (i.e., minimum-variance unbiased) cross-correlation estimators of $S_h$ is discussed in~\cref{sec:nonwhite}; see in particular equations~\eqref{e:Shcc_opt} and \eqref{e:sigma2_Shcc_opt_colored}.}
\begin{align}
&\hat P_I^{(0)} = D_{II}\quad{\rm for}\quad I=1,\cdots, \Nd\,,
\label{e:PI_quad}\\
&\hat S_h^{(0)}\equiv
\frac{1}{\Npairs}
\sum_{I=1}^{\Nd}\sum_{J>I}^{\Nd}
\frac{D_{IJ}}{\Gamma_{IJ}}\,,
\label{e:Sh_quad}
\end{align}
where $\Npairs\equiv \Nd(\Nd-1)/2$ is the number of distinct detector pairs.
The final outcome of the iteration gives 
{\it non-quadratic} estimators $\ML\theta_i$ of the parameters $\theta_i$.
An illustration of the difference between the quadratic and iterated ML estimators of the parameter $S_h$ for a single realization of data can be found in~\cref{fig:compare-3d} (grey and black dashed vertical lines).
But even for this simple 3-detector example, we were not able to find an analytic expression for the ML estimators.

In addition to the Whittle and Fisher likelihood functions, one can also write down the ``reduced'' Whittle likelihood approximation $\mathcal{L}^{\rm W,\, red}(S_h|\bar D;\bar P_1,\cdots,\bar P_{\Nd})$, see~\cref{e:LWR}.  
As discussed in~\cref{sec:2d}, the reduced Whittle likelihood is conditioned on the parameters $P_I$, assuming fixed, fiducial values $\bar P_I$ for these parameters.
Recall that the matrices $\bar P_{IJ}$ and $\bar D_{IJ}$ differ from $P_{IJ}$ and $D_{IJ}$ only by having $\bar P_I\equiv \bar P_{II}$ as their diagonal elements.

Finally, the ``reduced'' cross-correlation Gaussian likelihood approximation ${\cal L}^{\rm G,\,cc}(S_h|\bar D;\bar P_1,\cdots,\bar P_{\Nd})$ has exactly the same form as ${\cal L}^{\rm G,\,cc}(S_h|D_{12};\bar P_1, \bar P_2)$ given in~\cref{e:LGcc}, with $\hat S_{h,\, {\rm cc}}$ and $\sigma^2_{\hat S_{h,\, {\rm cc}}}$ given by
\begin{equation}
\hat S_{h,\,{\rm cc}}\equiv
\frac{1}{\Npairs}
\sum_{I<J}
\frac{D_{IJ}}{\Gamma_{IJ}}\,,
\qquad
\sigma^2_{\hat S_{h,\, {\rm cc}}}\equiv\frac{1}{2\Ns\Nf}\frac{1}{\Npairs^2}\sum_{I<J}
\frac{\bar{P}_I \bar{P}_J}{\Gamma_{IJ}^2}\,.
\label{e:Sh_sigma2Sh}
\end{equation}
As noted earlier, the above variance is strictly valid only in the weak-signal limit.

A comparison of the likelihood functions ${\cal L}^{\rm W}$,
${\cal L}^{\rm F}$,
${\cal L}^{\rm W,\, red}$,
${\cal L}^{\rm G,\, cc}$,
for the 3-detector, stationary, white signal, and white noise case, is given in~\cref{fig:compare-3d}.
As we did in~\cref{fig:compare-2d}, we plot the marginalized likelihoods as functions of the single GW parameter $S_h$, normalizing the distributions so that they correspond to posterior distributions for $S_h$ assuming flat priors for all the parameters.
The difference between the values of the quadratic and iterated ML estimators of $S_h$ is apparent for this particular realization of the data.
By construction, the peak of the cross-correlation Gaussian approximation using fiducial estimates $\bar P_I$ of the auto-power agrees with the quadratic cross-correlation estimator $\hat S_{h, {\rm cc}}$ and not $\hat S_h$, where the latter is obtained by iterating equations~\cref{e:iterate} for the full Whittle likelihood.
Nonetheless, since both $\hat S_{h,\, {\rm cc}}$ and $\hat S_h$ satisfy $\langle \hat S_{h,\, {\rm cc}}\rangle=S_h$ and $\langle \hat S_{h}\rangle=S_h$, there is no intrinsic bias in these analyses. 
However, as mentioned earlier, the reduced analyses will underestimate the uncertainty in the recovery of $S_h$, since they are conditioned on the autopower $\bar P_I$, and the Gaussian cross-correlation analysis explicitly uses a fiducial estimate of the variance, which is strictly only valid in the weak-signal regime.

\subsection{3-detectors, non-white case}
\label{sec:nonwhite}

\begin{figure}[t!]
    \centering
    \includegraphics[width=\textwidth]{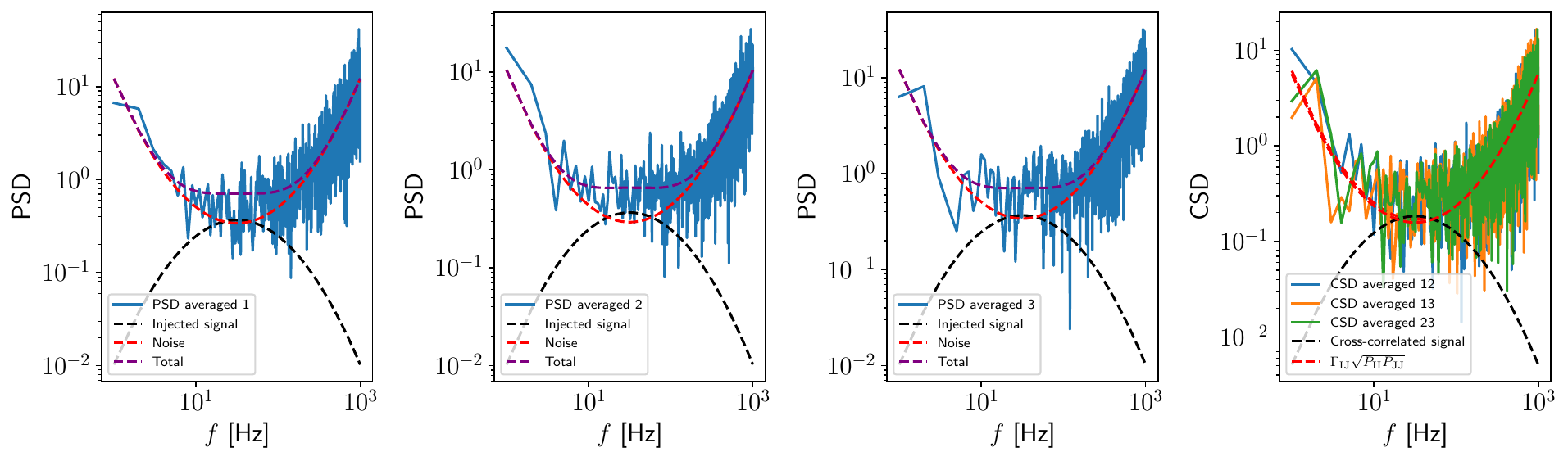}
    \caption{Power spectral density for every pair of detectors (channels) in a three-detector network. The three left panels show auto-spectra, while the rightmost panel displays all cross-spectra. Each colored line corresponds to one channel combination, with dashed lines indicating the injected signal model (black), noise (red), and their sum (purple). For this analysis we use $\Ns=3$ and $\Nf=10^3$.}
\label{fig:3channels_spectra_colored}
\end{figure}

We now extend the 3-detector example to allow for non-white signal and noise models.
Since the data are no longer white, we cannot naively compress the data $\tilde d_{I,k}^s$ over frequencies $f_k$ as we did in~\cref{e:DIJ}.
But, given our assumption of stationarity, we can still compress over segments:
\begin{equation}
D_{IJ,k}\equiv \frac{1}{\Ns}\sum_s \frac{2}{T}{\rm Re}[\tilde d_{I,k}^s \tilde d_{J,k}^{s*}]\,.
\label{e:D-nonwhite}
\end{equation}
For simplicity, we will assume that both the signal and noise power spectra $S_h(f_k)$ and $P_{n_I}(f_k)$ have known shapes $H(f_k)$ and $N_I(f_k)$, i.e., 
\begin{equation}
S_h(f_k) = S_h\, H(f_k)\,,
\quad
P_{n_I}(f_k) = P_{n_I}\, N_I(f_k)\,,
\end{equation}
where $H(f_k)$ and $N_I(f_k)$ are normalized to unity at some reference frequency. For illustrative purposes, in our analysis, we modeled both the signal and noise spectra as simple parabolic functions of $x\equiv\log(f/f_*)$ with $f_*$ = $\sqrt{f_{\min}f_{\max}}$ for signal and noise spectra. In~\cref{fig:3channels_spectra_colored} we plot the resulting power spectral densities for these injected signal and noise models.
The parameters that we will need to infer for this example are then the overall amplitudes of the power spectra, $\theta\equiv (P_{n_1}, P_{n_2}, P_{n_3}, S_h)$.

\begin{figure}
    \centering
    \includegraphics[width=0.6\textwidth]{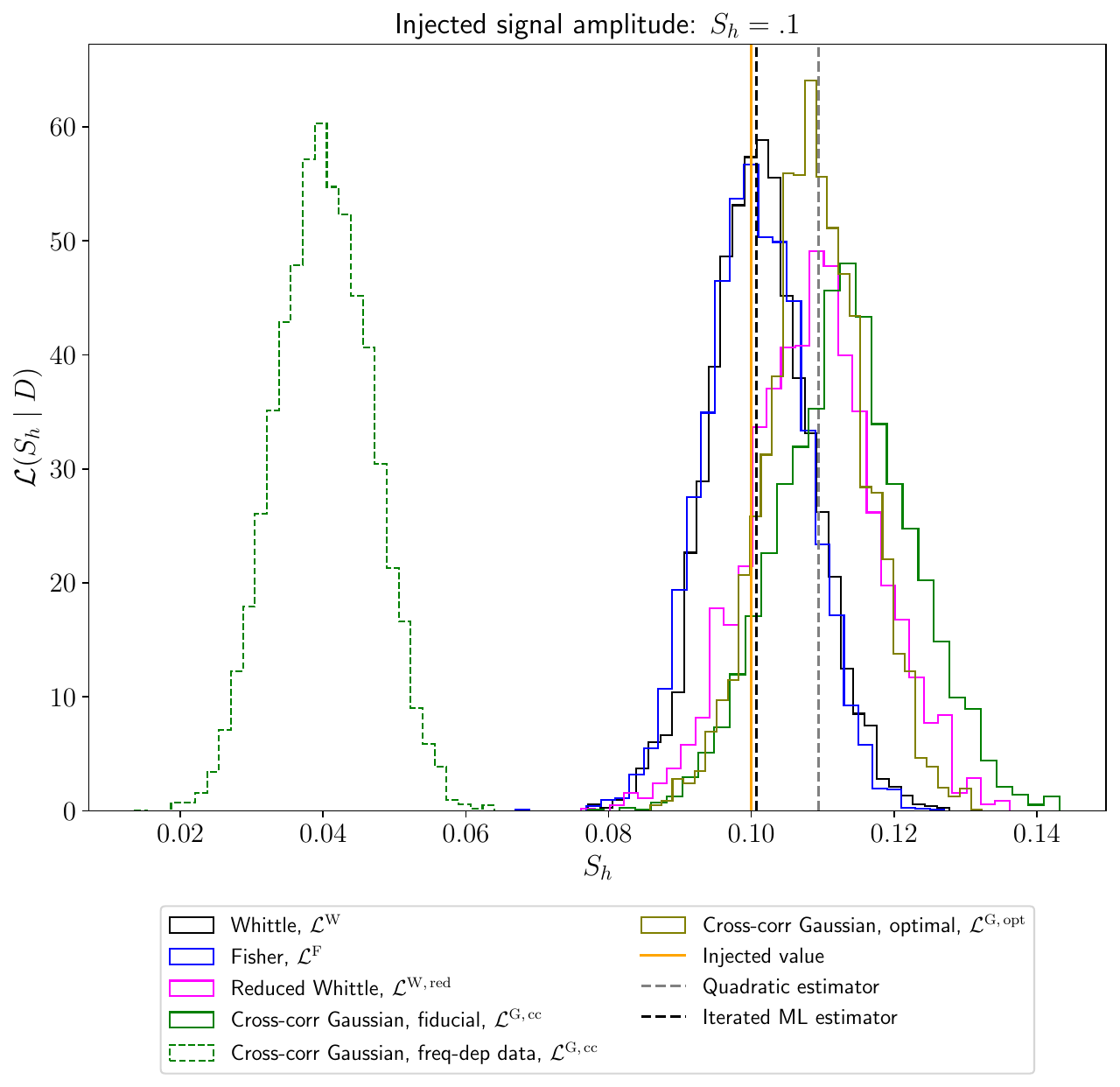}
    \caption{Comparison of the posterior distributions for $S_h$ using the Whittle, Fisher, reduced Whittle, and cross-correlation approximate likelihood functions for the three-detector, colored noise and signal example. For this analysis, we have injected a signal with $S_{h}= 0.1$, noise amplitude $P_{n_1} = 1.25$, $P_{n_2} = 1.08$ and $P_{n_3} = 1.25$ for each detector, and we have set $\Ns = 3$ and $N_f = 1000$.}
    \label{fig:3channels_constraints_colored}
\end{figure}

If we assume that the noise is uncorrelated across detectors, then the expected value of 
$D_{IJ,k}$ is 
\begin{align}
&\langle D_{IJ,k}\rangle = P_{IJ,k}\,,
\label{e:DIJk-nonwhite}
\quad{\rm where}\quad
P_{IJ,k}\equiv \delta_{IJ} P_{n_I} N_{I,k} +  S_h\, \Gamma_{IJ} H_k\,.
\end{align}
In terms of the matrices $D_{k}$ and $P_k$ (with components $D_{IJ,k}$ and $P_{IJ,k}$ defined above), the four likelihood functions we defined in previous subsections become
\begin{widetext}
\begin{align}
&-2 \ln \mathcal{L}^{\rm W}
(\theta\vert D_k) =2\Ns\sum_k\left[\ln \det(P_k)
+ {\rm Tr}(P_k^{-1}D_k)\right] \,,
\label{e:lnLWishart}
\\
&-2 \ln \mathcal{L}^{\rm F}
(\theta\vert D_k) = \sum_{i,j}(\hat \theta_i-\theta_i)\hat F_{ij}(\hat \theta_j - \theta_j) \,,
\label{e:lnLFisher}
\\
&-2 \ln \mathcal{L}^{\rm W,\, red}
(S_h\vert \bar D_k;\bar P_{11,k},\cdots) =2\Ns\sum_k\left[\ln \det(\bar P_k)
+ {\rm Tr}(\bar P_k^{-1}\bar D_k)\right] \,,
\label{e:lnLReduced}
\\
&-2 \ln \mathcal{L}^{\rm G,\, {\rm cc}}
(S_h\vert \bar D_k;\bar P_{11,k},\cdots) =
\sum_k \frac{(\hat S_{h,\, k} - S_h)^2}{\sigma^2_{\hat S_{h,\, k}}} \,,
\label{e:lnLGaussian}
\end{align}
which are similar in form to what we had in previous subsections.
But now the data combination and power spectra depend on frequency, and there is a summation over the discrete frequencies.
The expressions for $\bar D_k$ and $\bar P_k$ are also similar to what we had earlier; namely, we replace the diagonal elements of $D_{IJ,k}$ and $P_{IJ,k}$ with estimates of the total auto-power $\bar P_{II,k}$.  
This replacement means that, for the reduced analyses, we no longer need to infer the amplitudes of the noise power spectra $P_{n_I}$.
In addition, for the reduced Gaussian cross-correlation likelihood, we have
\begin{equation}
\begin{aligned}
\hat S_{h,\,k}\equiv
\frac{1}{\Npairs}
\sum_{I<J}
\frac{D_{IJ,k}}{\Gamma_{IJ}}\,,
\qquad
\sigma^2_{\hat S_{h,\, k}}\equiv\frac{1}{2\Ns}\frac{1}{\Npairs^2}\sum_{I<J}
\frac{\bar{P}_{II,k} \bar{P}_{JJ,k}}{\Gamma_{IJ}^2}\,.
\label{e:Shk+sigma2Shk}
\end{aligned}
\end{equation}
which are similar to~\cref{e:Sh_sigma2Sh}, but using the frequency-dependent quadratic data combinations $D_{IJ,k}$ for $I\ne J$ and fiducial auto-power estimates $\bar P_{II,k}$. Notice that depending on the choice for $\bar{P}_{II,k}$, i.e., some fiducial estimate smoothed over neighboring frequency bins vs a purely frequency-by-frequency data-driven estimate, the reduced likelihood $\mathcal{L}^{\rm W,red}$ and the cross-correlation likelihood $\mathcal{L}^{\rm G,cc}$ can return unbiased/biased results. An explicit example of this behavior is shown explicitly in~\cref{fig:3channels_constraints_colored} for $\mathcal{L}^{\rm G,cc}$ (note the difference between the green and dashed-green distribution).
\end{widetext}

For the reduced Gaussian cross-correlation likelihood, it is also possible to further compress the data over frequency, using information about the expected shape of the signal and noise models.
In particular, we can form an {\it optimal} (i.e., minimum-variance,%
\footnote{This estimator of $S_h$ has minimum variance when restricted to the class of quadratic cross-correlation estimators. 
The unrestricted minimum-variance, unbiased estimator of $S_h$ comes from the full Whittle likelihood (it saturates the Cramer-Rao bound~\cite{kay1993estimation}), using information from auto-correlations.
As mentioned previously, the Whittle likelihood ML estimator is non-quadratic and requires iteration for detector networks having three or more detectors.}
unbiased) estimate of $S_h$, which is summed over frequencies and distinct detector pairs using information about the mean and covariance of $D_{IJ,k}$, which we calculate below.

But, to first understand optimal filtering in the context of a simpler example, suppose we have $N$ samples of random data $x_i$ with mean and covariance
\begin{equation}
\begin{aligned}
\langle x_i\rangle = A\, \Gamma_i\,,
\quad
\langle x_i x_j\rangle - \langle x_i\rangle\langle x_j\rangle= C_{ij}\,.
\end{aligned}
\end{equation}
We note that the $x_i$ do not have to be Gaussian distributed.
Let us further assume that the mean ``shape'' $\Gamma_i$ and covariance matrix $C_{ij}$ are known a~priori, or that they can be estimated from neighboring data segments or via some other means, e.g.,  making spline fits to the data.
Then the optimal quadratic estimator of the amplitude $A$ is given by
\begin{equation}
\hat A = \frac{\sum_{i,j}\Gamma_i (C^{-1})_{ij} x_j}{\sum_{k,l}\Gamma_k (C^{-1})_{kl}\Gamma_l}\,,
\label{e:Ahat}
\end{equation}
which can be thought of as an ``inverse-noise weighted matched filter'' statistic, normalized such that $\langle \hat A\rangle=A$.
In practice, the inverse noise weighting ``whitens'' the data.
The variance of the estimator $\hat A$ is given by 
\begin{equation}
\sigma^2_{\hat A}= \frac{1}{\sum_{k,l}\Gamma_k (C^{-1})_{kl}\Gamma_l}\,.
\end{equation}
Note that~\cref{e:Ahat} reduces to the standard $1/\sigma^2_i$ weighting if the covariance matrix $C$ is diagonal.

\begin{widetext}

We already calculated the mean value of $D_{IJ,k}$ in~\cref{e:DIJk-nonwhite}.
The covariance matrix of  $D_{IJ,k}$ is given by
\begin{equation}
\begin{aligned}
C_{IJk,KLk'}
&\equiv\langle D_{IJ,k} D_{KL,k'}\rangle-\langle D_{IJ,k}\rangle\langle D_{KL,k'}\rangle
\\
&=\frac{1}{2\Ns}\delta_{kk'}
\left(P_{IK,k} P_{JL,k} + P_{IL,k}P_{JK,k}\right)\,,
\\
&\equiv C_{IJ,KL}\,\delta_{kk'}\,,
\end{aligned}
\end{equation}
where we used Isserlis's theorem~\cite{isserlis:1918} to expand the 4th-order expectation values of products of the Gaussian data $\tilde d_{I,k}^s$ which enter in $D_{IJ,k}$.

Since the noise is uncorrelated, we can construct an optimal quadratic estimator for the amplitude $S_h$ of the GW power spectrum using only the cross-correlated data combinations $D_{IJ,k}$ where $I\ne J$.
This reduces the range of indices for the covariance matrix $C_{IJ,KL}$.
Explicitly, for the case of 3 detectors, we can write
\begin{equation}
C_{IJk,KLk'}
= \frac{1}{2\Ns}\delta_{kk'}
\left[\begin{matrix}
P_{11,k} P_{22,k} + P_{12,k}^2 &
P_{11,k} P_{23,k} + P_{12,k} P_{13,k} &
P_{12,k} P_{23,k} + P_{13,k} P_{22,k} 
\\
P_{11,k} P_{23,k} + P_{12,k} P_{13,k} &
P_{11,k} P_{33,k} + P_{13,k}^2 &
P_{12,k} P_{33,k} + P_{13,k} P_{23,k} 
\\
P_{12,k} P_{23,k} + P_{13,k} P_{22,k} &
P_{12,k} P_{33,k} + P_{13,k} P_{23,k} &
P_{22,k} P_{33,k} + P_{23,k}^2 
\end{matrix}
\right]\,.
\end{equation}
We will further restrict attention to the case of a ``reduced analysis'' where we condition on the fiducial estimates of $P_{II,k}$ and also assume that the GWB is weak compared to the noise.
Then
\begin{equation}
\begin{aligned}
\bar C_{IJk,KLk'}
&= \frac{1}{2\Ns}\delta_{kk'}
\left[\begin{matrix}
\bar P_{11,k} \bar P_{22,k} & 0 & 0 
\\
0 & \bar P_{11,k} \bar P_{33,k} & 0
\\
0 & 0 &  \bar P_{22,k} \bar P_{33,k} 
\end{matrix}
\right]
=  \frac{1}{2\Ns}\delta_{kk'}\delta_{IK}\delta_{JL}\bar P_{II,k}\bar P_{JJ,k}\,,
\end{aligned}
\end{equation}
which allows us to drop all cross-spectra terms $P_{IJ,k}$ (for $I\ne J$) relative to the auto-power spectra $P_{II,k}$.
We put a bar over the covariance matrix to indicate it is conditioned on the power spectrum estimates.
\end{widetext}

In terms of this reduced covariance matrix, the optimal cross-correlation estimator of $S_h$ is given by the data combination 
\begin{equation}
\hat S_{h,\, {\rm opt}}
={\cal N}\left[\sum_{I<J}\sum_k\frac{\Gamma_{IJ} H_k D_{IJ,k}}{\bar P_{II,k}\bar P_{JJ,k}}\right]  \,,
\label{e:Shcc_opt}
\end{equation}
with variance 
\begin{equation}
\sigma^2_{\hat S_{h,\, {\rm opt}}}
\equiv\frac{1}{2\Ns}{\cal N}
\equiv\frac{1}{2\Ns}
\left[\sum_{I<J}\sum_k\frac{\Gamma_{IJ}^2 H_k^2}{\bar P_{II,k}\bar P_{JJ,k}}\right]^{-1}\,,
\label{e:sigma2_Shcc_opt_colored}
\end{equation}
which is valid in the weak-signal limit.
In terms of this optimal estimator, 
the reduced Gaussian cross-correlation likelihood function, denoted $\cal L^{\rm G,\, opt}$ is then
\begin{equation}
-2\ln {\cal L}^{\rm G,\,opt}(S_h|\bar D_k;\bar{P}_{11,k}, \cdots) =\frac{(\hat S_{h,\,{\rm opt}} - S_h)^2}{\sigma^2_{\hat S_{h,\,{\rm opt}}}}\,,
\label{e:LGcc_colored}
\end{equation}
with $\hat S_{h,\,{\rm opt}}$ and $\sigma^2_{\hat S_{h,\,{\rm opt}}}$ given above.\\

In~\cref{fig:3channels_constraints_colored} we plot a comparison of the likelihood functions ${\cal L}^{\rm W}$,
${\cal L}^{\rm F}$,
${\cal L}^{\rm W,\, red}$,
${\cal L}^{\rm G,\, cc}$, and
${\cal L}^{\rm G,\, opt}$,
for the 3-detector, stationary, non-white signal, and non-white noise case.  We have injected a signal with $S_{h} = 0.1$ and noise amplitude for each detector sampled from a normal distribution $\mathcal{N}(0, 1.25)$.
As before, we plot the marginalized likelihoods as functions of the single GW parameter $S_h$, normalizing the distributions so that they correspond to posterior distributions for $S_h$ assuming flat priors for all the parameters.
For the reduced analyses, we obtain unbiased/biased results depending on whether $\bar{P}_{II,k}$ is given by some fiducial estimate smoothed over neighboring frequency bins or uses a purely frequency-by-frequency data-driven estimate.  This is illustrated by the green and dashed-green distributions for the ${\cal L}^{\rm G,\, cc}$ likelihood analyses in~\cref{fig:3channels_constraints_colored}. 

We conclude this section by referring the reader to~\cref{sec:appendix_tests}, where we perform a series of tests to assess the robustness of the different likelihoods in different signal, frequencies, and segments regimes. In particular, we show the posterior distribution in the strong signal regime and varying the number of frequencies/segments, and we produce some P-P plots to study the coverage accuracy of the different likelihoods.

\section{Realistic GW detectors}\label{sec:realistic_detectors}

In the data analysis of current and future detectors, the choice of the number of time segments used in the data processing plays a crucial role in determining the accuracy and reliability of the likelihood function employed to detect and characterize the SGWB. 
Segmenting the continuous data stream into $\Ns$ segments ensures the validity of the stationarity and Gaussianity assumptions within each segment, significantly simplifying the analysis and eventually avoiding biases in the parameter estimation.
In this section, we estimate the expected fractional statistical uncertainty ${\cal S}$ with present and future GW detectors and compare it to the expected bias ${\cal B}$ obtained through the Gaussian-based estimator defined in~\cref{e:bias}. The main results of this section are summarized in~\cref{tab:configs_summary}.

A technical point to mention is that, typically, the analysis uses the full GW-sensitive band (up to a given $f_{\rm max}$) to estimate signal and noise simultaneously. In practice, different analyses focus on narrower frequency (or time–frequency) regions, where the target signal is most probable (i.e., where the signal-to-noise ratio is more likely to be larger) or stationary. %
As a result, the true effective bandwidth used to estimate signal and noise is narrower than the entire detector range in most searches, which can alter both statistical uncertainties and systematic biases compared to the idealized, full-band estimates. To try to address this point, we also provide uncertainties using the FIM derived as in~\cref{e:F}, i.e., taking derivatives of the detector network Whittle likelihood with respect to the model parameters and weighting them by the inverse of the signal and noise PSDs. 
For reference, we inject a flat, i.e., frequency‑independent, SGWB with amplitude $\Omega_{\rm GW} = 5.8 \times 10^{-9}$, corresponding to the 95\% credible upper limit reported by the LVK collaboration~\cite{KAGRA:2021kbb}. 
These choices maximize the amplitude of the possible signal in the ground-based detector's frequency band, while indicating a representative benchmark for experiments in different frequency ranges, and thus give an optimistic estimate of the sensitivity. Clearly, signals with a lower amplitude would correspond to larger parameter uncertainties.
The noise curves adopted for the individual detectors are the design‑sensitivity strain power spectral density $P_n(f)$ for Advanced LIGO~\cite{Barsotti:18}, the ET‑D noise configuration for the Einstein Telescope~\cite{Branchesi:2023mws}, and the noise‑equivalent strain spectral density $S_n(f)$ for the LISA detector~\cite{LISA:2017pwj}.
For current and future PTA datasets, we model pulsar noises based on the currently observed pulsars and generate future catalogs with similar properties (see~\cite{Babak:2024yhu} for more details).

\subsection{Possible biases in isotropic searches}

Ground-based detectors, such as the current LVK network and the future ET and CE detectors, operate in the frequency band between 1 Hz and several kHz. The LVK detectors are interferometers with arm lengths of a few km designed to detect GWs produced by compact binary mergers (mainly black holes and neutron stars), supernovae, and, at design sensitivity, also the SGWB of astrophysical origin~\cite{KAGRA:2021kbb}. The current event rate is approximately one merger every 2 to 3 days~\cite{KAGRA:2021vkt}. LVK is now in its fourth run (O4), which is supposed to last until fall 2025 for a total of $T_{\rm obs} \simeq 18$ months (excluding commissioning breaks). 

Typical data segments in LVK analysis last from 4 seconds to several (10) minutes, depending on the nature of the targeted signal, where the noise is (roughly speaking) stationary. Assuming $f_{\rm max} = 10^3~{\rm Hz}$ and $T_{\rm obs} = 18$ months, we report in~\cref{tab:configs_summary} the values for the statistical uncertainties $\mathcal{S}$, bias $\mathcal{B}$, and relative uncertainty on the signal amplitude for two typical segmentation time values. We notice that with 4s segments, the systematic bias is below the statistical uncertainty, and more than three orders of magnitude below the FIM bound. Therefore, the measurement is statistics‑limited. With $T = 10~{\rm min}$, the bias increases but remains well below the statistical uncertainty estimated with the FIM. 

ET~\cite{Abac:2025saz} is a proposed third-generation GW detector with a triangular configuration of three 10-kilometer interferometers (see~\cite{Branchesi:2023mws} for other possible configurations) situated underground to minimize the impact of seismic noise at low frequency. Similarly, CE will be a third-generation L-shaped interferometer with 40-km arm length~\cite{Evans:2023euw}. They will have at least one order of magnitude better sensitivity than second-generation detectors and will extend the observable frequency range down to 1 Hz, which will allow for the detection of signals from a wide variety of astrophysical sources, including the SGWB. ET and CE are expected to significantly enhance detection rates of compact binary coalescences due to their advanced sensitivity. Current forecasts estimate that ET could detect approximately $10^5$ BBH mergers per year, along with a similar number of BNS mergers~\cite{Abac:2025saz,Pieroni:2022bbh}. This large increase in event rates, compared to current detectors, will provide a more comprehensive understanding of the population of compact binary systems~\cite{ET:2019dnz, Abac:2025saz}. 

Even if a proper segmentation strategy is not yet defined for ET/CE, as long as all the noise contributions are not well characterized,  assuming a similar duration as LVK (from a few seconds up to several minutes) will lead to a large number of $\Ns$ per year. 
The choice of such a large number of shorter segments helps to ensure that the noise can be treated as stationary and Gaussian within each segment. For ET and CE, a 4-sec segmentation yields a bias that is two orders of magnitude smaller than the statistical uncertainty and is entirely negligible compared to the FIM bound. However, when 10-min segments are considered, the bias exceeds the statistical error and becomes a dominant contribution to the total uncertainty, indicating a possible transition to a bias-limited regime.

\begin{table*}[t]
\centering
\renewcommand{\arraystretch}{1.4}
\begin{tabular}{|l|c|c|c|c|c|c|}
\hline
Detector\;config. (seg duration $T$) & $T_{\rm obs}$ & $\Ns$ & $N_f$ & $\cal S$ & $\cal B$ & $\cal S \ \text{from} \ {\rm FIM}$ \\
\hline
\multicolumn{7}{|c|}{Ground-based detectors} \\
\hline
LVK\,(4\;s) & \multirow{2}*{18\,mo}  & $1.2\times10^{7}$ & $4.0\times10^{3}$ & \multirow{2}*{$4.6\times10^{-6}$} & $8.0\times10^{-8}$ &
\multirow{2}*{$1.4 \times 10^{-2}$}  \\
\cline{1-1}\cline{3-4}\cline{6-6}
LVK\,(10\,m) &  & $7.8\times10^{4}$ & $6.0\times10^{5}$ &  & $1.3\times10^{-5}$ &
\\
\hline
ET (CE)\,(4\;s)             & \multirow{2}*{12\,mo} & $7.9\times10^{6}$ & $4.0\times10^{3}$ & \multirow{2}*{$5.6\times10^{-6}$} & $1.3\times10^{-7}$ &
\multirow{2}*{$4.6 (2.5)  \times 10^{-5}$ }  \\
\cline{1-1}\cline{3-4}\cline{6-6}
ET (CE)\,(10\,m)            &   & $5.3\times10^{4}$ & $6.0\times10^{5}$ &   & $1.9\times10^{-5}$ &
\\
\hline
\multicolumn{7}{|c|}{Space-based detectors} \\
\hline
LISA\,(11.5\,d)           & 4\,yr  & $1.3\times10^{2}$ & $5.0\times10^{5}$ & $1.3\times10^{-4}$ & $8.0\times10^{-3}$ &
$4.2 \times 10^{-4}$ \\
\hline
Taiji\, (11.5\,d)             & 6\,yr  & $1.9\times10^{2}$ & $5.0\times10^{5}$ & $1.0\times10^{-4}$ & $5.2\times10^{-3}$ &
$2.8 \times 10^{-4}$ \\
\hline
\multicolumn{7}{|c|}{Galactic-scale detectors} \\
\hline
PTAs current\;(single segment)    & 15\,yr & $1$               & $\sim10$      &
$4 \times 10^{-2}$ & -- & $1.3 \times 10^{-1}$  \\
\hline
PTAs future\;(single segment)    & 15\,yr & $1$               & $\sim10^{2}$      & $7 \times 10^{-3}$ & -- & $3 \times 10^{-2}$\\
\hline
\end{tabular}
\caption{
Representative number of segments \(\Ns\), independent frequency bins \(N_{f}\), minimal statistical uncertainty \({\cal S}\), and potential systematic bias \({\cal B}\) for various detector configurations. In the last column, we report the relative uncertainty on the signal amplitude, estimated using the FIM, assuming a flat SGWB injected with an amplitude matching the current LVK upper bound~\cite{KAGRA:2021kbb}, i.e., $\Omega_{\rm GW} = 5.8 \times 10^{-9}$. 
Ground-based figures assume \(f_{\max}=10^{3}\,\mathrm{Hz}\); space-based and PTA entries follow the specifications discussed in the text.}
\label{tab:configs_summary}
\end{table*}

LISA consists of a constellation of three spacecraft in an equilateral configuration, with arm lengths of 2.5 million kilometers~\cite{LISA:2017pwj}. This configuration is designed to detect low-frequency GW signals,  such as those emitted by supermassive black hole binaries (SMBHBs) and extreme mass-ratio inspirals (EMRIs), which cannot be observed by ground-based detectors. LISA will operate in the milli-Hz frequency regime (from $3 \times 10^{-5}$ Hz up to 0.5 Hz) for a baseline mission of at least 4.5 years, potentially extending up to 10 years. Motivated by the cadence of scheduled satellite operations, the data are commonly divided into relatively long segments. LISA will have an effective observation time of 
$T_{\rm obs} = 4$~years, and the segment duration will be $T = 11.5$ days each. This corresponds to a frequency resolution $\Delta f$  = $10^{-6}$ Hz. In this way, we get $N_{\rm s} \simeq 126$ and $N_f \simeq 5 \times 10^5$. 
In~\cref{tab:configs_summary}, we report the values for the bias ${\cal B}$, the statistical uncertainty $ {\cal S}$, and the FIM estimates for the LISA detector.  We notice that the bias $ {\cal B}$ can exceed both the statistical  $ {\cal S}$ and FIM uncertainty. 

Other forthcoming space-based GW observatories, such as Taiji~\cite{Ruan:2018tsw} or TianQin~\cite{TianQin:2015yph}—which are also designed to work in the milli-Hz band with comparable baselines—will face similar challenges. They will likely adopt analogous segmentation strategies and must likewise account for biases that can exceed statistical uncertainties in their data analysis. In~\cref{tab:configs_summary}, we compute the expected uncertainties also for Taiji, considering $T_{\rm obs} = 6$ years, $T = 11.5$ days, and the same frequency resolution as LISA.

Finally, we consider the case of Pulsar Timing Arrays (PTAs). PTAs represent an international effort of four collaborations—NANOGrav, EPTA, PPTA, and CPTA—that monitor the pulse arrival times of ultra-stable millisecond pulsars across the sky~\cite{NANOGrav:2023gor,NANOGrav:2023hde, EPTA:2023fyk, Reardon:2023gzh, Xu:2023wog}. By analyzing timing residuals accumulated over more than a decade (15 yrs), they have independently measured the distinctive quadrupolar correlation pattern predicted by Hellings and Downs—a result interpreted as the first detection of a nanohertz-band SGWB. However, the physical origin of such a background remains uncertain: it could arise from a population of inspiraling supermassive black-hole binaries or from cosmological sources, such as cosmic strings, first-order phase transitions, or scalar-induced gravitational waves~\cite{NANOGrav:2023hvm,EPTA:2023xxk,Figueroa:2023zhu,Ellis:2023oxs}. 

Because the effective observation time is set by the entire data span, PTAs operate with essentially a single Fourier segment ($\Ns\sim 1$). So the suppression of statistical noise is absent; therefore, the experiment is highly sensitive to biases, as can be seen in~\cref{tab:configs_summary}, where we compute the expected uncertainties for current and future PTA detector configurations. 
For the future PTA configuration, we assumed $T_{\rm obs} = 15~{\rm yr}$ observations of $N_p = 200$ pulsars with an SKA-like noise configuration (see, e.g., for more detail~\cite{Babak:2024yhu}). For the PTA result only, we estimate the fractional statistical uncertainties using $1/\sqrt{\Nf \Ns N_p^\alpha}$, including the scaling with the number of pulsars, which here is controlled by the parameter $\alpha$. Depending on whether cross-correlations or auto-correlations carry more information, $\alpha$ can smoothly interpolate between $\alpha=2$ and $\alpha=1$ respectively. Notice that, upon replacing $N_p$ with $N_{\rm det}$, a similar scaling would also apply when combining the output of several interferometers.  

\subsection{A remark on anisotropic searches}

Anisotropic or directional SGWB searches using LVK data use a data folding technique to exploit Earth’s rotation~\cite{Ain:2015lea,Ain:2018zvo,KAGRA:2021mth}. Instead of treating each of the $\Ns$  segments separately, the cross-correlation data are compressed into one sidereal day (23 h 56 m) of effective time. This is done by ``folding'' the time-series: segmenting data in 192s chunks aligned to sidereal time and stacking them modulo 1 sidereal day. Following this procedure, the summations over number of segments can be decomposed as $\sum_s \rightarrow \sum_{i_{\text{day}}} \sum_{t_s},$ where $i_{\text{day}}$ runs over all sidereal days included in the data set, and $t_s$ spans a single sidereal day. 
In this way, each sidereal day contains $\simeq 448$ segments of 192 s (since $24_{\text{sid}}{\rm h}\approx86164{\rm s}$), which repeat every day. Thus, the data are mapped onto these $\simeq 448$ sidereal time bins. 

For instance, in the O1+O2 runs, the anisotropic search folded 129 days of data into one sidereal day, i.e., each sidereal-time bin averages over $\approx 129$ independent segments (one from each day). Similarly, for the O3 anisotropic analysis, 169 days of data were folded into 1 sidereal day. So,  each of the sidereal bins contains the averaged cross-power from 169 different days at that sidereal time. 

Given these remarks, it might appear that the effective number of independent time segments in a folded anisotropic search is of order only a few hundred (the number of sidereal bins) per baseline, much lower than in the isotropic case, which compresses all segments in one large average (losing sky-direction information). However, the folded anisotropy search avoids this issue by performing an additional operation, assuming a stationary distribution of GW power on the sky. Given the data folded onto the sidereal day, the time and direction-dependent (meaning no average over sky-directions) overlap reduction function is deconvolved from the data, allowing compression over the time bins of the sidereal day. At the end of this procedure, anisotropic searches effectively compress over the same number of segments as the isotropic search, which might preserve approximate Gaussianity at the level of accuracy set by the statistical uncertainty.

\section{Conclusions and outlook}\label{sec:conclusions}

We have presented a detailed, end-to-end investigation of the Bayesian framework underpinning SGWB searches. We have tested the validity of using the standard Gaussian likelihood employed by current LVK searches to model the cross-correlation statistics. 

Beginning with the exact Whittle likelihood in the simplest setting of a single detector and a single frequency bin, we have shown analytically that replacing the true likelihood with some approximated version can introduce a bias relative to the maximum-likelihood estimator. Then, we have extended these toy models to multiple (i.e., two and three) detectors for both white and realistic colored noise, and we have quantified how the bias $\mathcal{B}$ and the statistical uncertainty $\mathcal{S}$ depend on segment length $T$, total observation time, and bandwidth. 

For second-generation ground-based detectors (LVK) and next-generation detectors (ET/CE), we find that very short segments $\mathcal{O}(1)$ s yield $\mathcal{B}\ll\mathcal{S}$, preserving a ``statistics-limited'' regime.  However, pushing $T$ into the $\mathcal{O}(10)$ minutes regime can bring $\mathcal{B}$ to grow larger than $\mathcal{S}$, inducing a significant bias in the estimation of the SGWB  energy density. 
For space-based missions (LISA/Taiji), which typically divide their multi-year observation into  $\mathcal{O}(10^2)$ segments of roughly $\sim 11\,\mathrm{\rm days}$ length, such a segmentation can lead to sizable bias, which typically exceeds the statistical uncertainty. 
Finally, for PTA $\Ns\sim1$, which places them in the bias-limited regime. We discussed the impact of our investigations for anisotropic SGWB searches, We concluded that, since sidereal-day folding reorganizes, but does not reduce, the effective sample count, the results derived for isotropic searches continue to be valid.

To better quantify the uncertainties, we have developed two classes of modified likelihoods.  The ``reduced'' Whittle likelihood conditioned on fiducial data-driven autocorrelation spectra, and the ``optimal-filter'' one, which applies inverse-noise weighting across frequency and detector pairs to minimize variance without the problem of extra bias.  Both methods recover unbiased maximum-likelihood estimates and could be easily integrated into existing stochastic pipelines.

As the detector sensitivities improve and SGWB detections become affordable, it will be essential to incorporate these refined likelihoods into the future LVK analyses, LISA/ET/CE data challenge, and PTA pipelines. Combining these likelihoods with optimized segment choices and data-compression strategies can ensure SGWB detection and robust, unbiased parameter reconstruction. Including non-Gaussian noise transients, frequency-frequency correlations, and/or time-varying noise covariances are interesting possibilities for further generalizations. Moreover, realistic mock data, including resolved and stochastic signals together with additional non-stationary features, would be important to validate our estimator's performance. We leave all these investigations for future work.

\begin{acknowledgments}
We thank James Alvey for his very valuable comments on a nearly final version of this manuscript. We also thank Jessica Lawrence and Kevin Turbang for providing the ``dictionary'' plot in~\cref{fig:pp-plot-dictionary}. MP acknowledges the hospitality of Imperial
College London, which provided office space during some
parts of this project. The work of MP is supported by the Comunidad de Madrid under the Programa de Atracción de Talento Investigador with number 2024-T1TEC-31343.
JDR acknowledges financial support from the National Science Foundation (NSF) Grant No. PHY-2207270 and start-up funds from the University of Texas Rio Grande Valley.
\end{acknowledgments}

\appendix

\section{Further tests}
\label{sec:appendix_tests}
\begin{figure}
    \centering
    \includegraphics[width=0.46\textwidth]{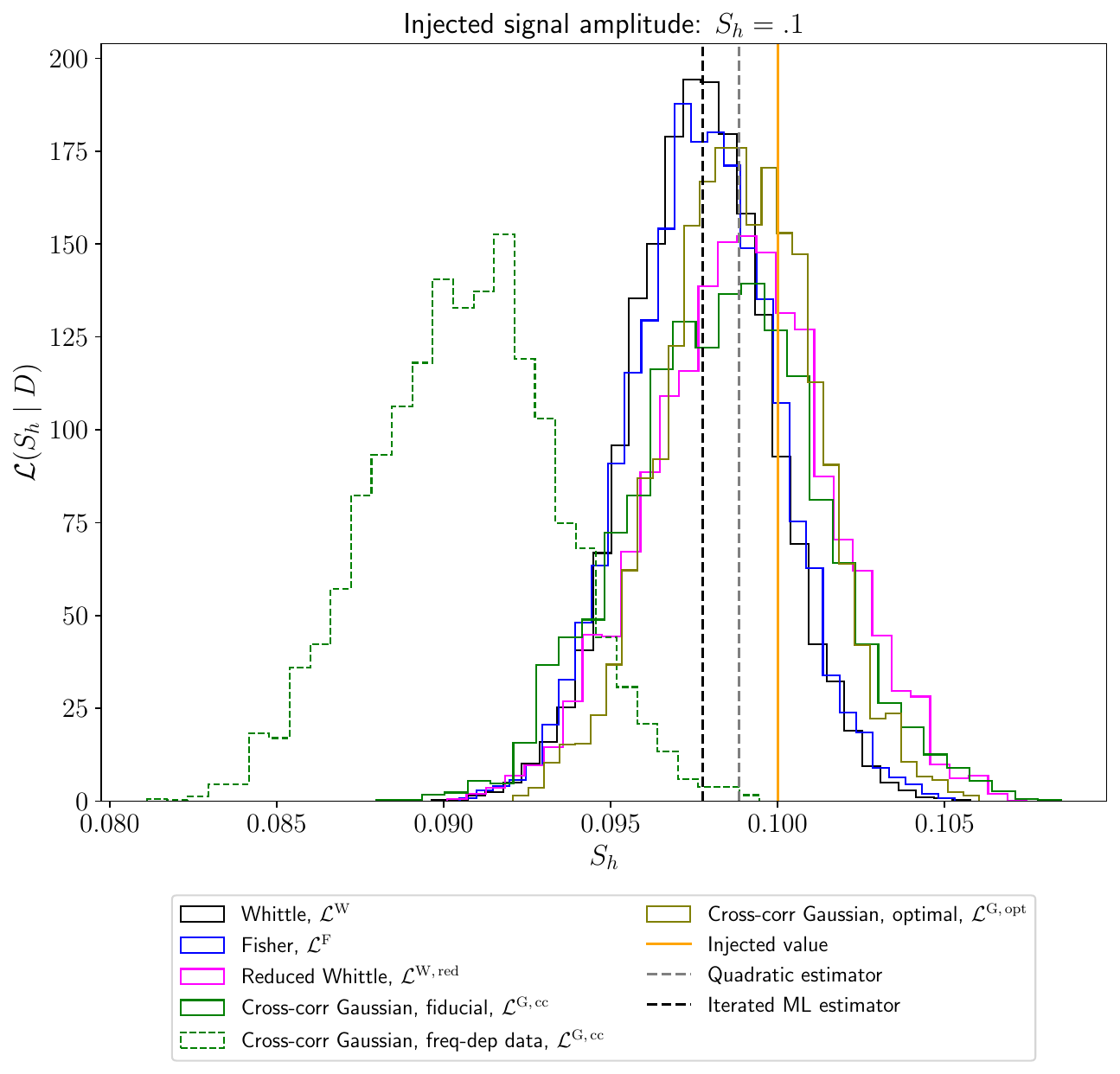}
    \includegraphics[width=0.46\textwidth]{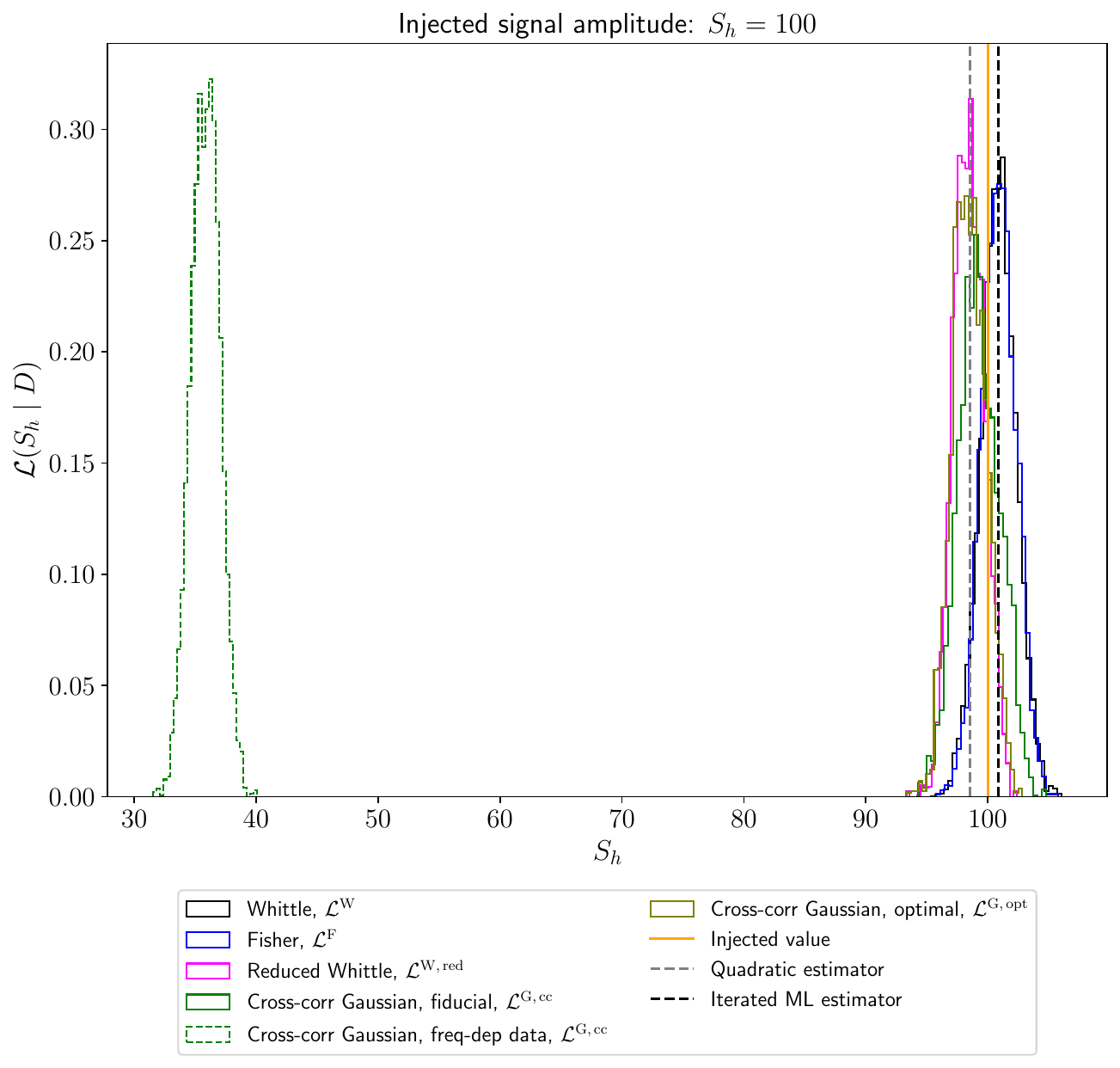}
    \caption{Left: 
    Similar to~\cref{fig:3channels_constraints_colored}, with an injected signal with $S_{h}= 0.1$, noise amplitude $P_{n_1} = 1.25$, $P_{n_2} = 1.08$ and $P_{n_3} = 1.25$ for each detector, but increasing the number of segments to $\Ns = 32$.
    Right: Similar to~\cref{fig:3channels_constraints_colored}, with an injected signal with $S_{h}= 10^2$, noise amplitude $P_{n_1} = 1.25$, $P_{n_2} = 1.08$ and $P_{n_3} = 1.25$ for each detector, and number of segments $\Ns = 3$.}
    \label{fig:3channels_constraints_colored_app}
\end{figure}

In this appendix, we present a series of additional tests for addressing the robustness of the different likelihoods and to study their behaviour in different signal-to-noise regimes. In~\cref{fig:3channels_constraints_colored_app}, we present a comparison of the posterior distributions for the different likelihoods increasing the number of segments to $\Ns = 32$ (left plot) or injecting a ``very strong'' signal, $S_h = 100$ (right plot), to compare with the behaviour of the posteriors in the case of a low number of segments  $\Ns = 3$ and ``intermediate'' signal (i.e., comparable with noise) $S_h = 0.1$, presented in the main text in~\cref{fig:3channels_constraints_colored}. 
In the left plot, we notice that, as $\Ns$ increases, the uncertainties become smaller and the bias decreases relative to the uncertainty. On the other hand, as shown in the right plot, by increasing the signal amplitude, the relative uncertainty (i.e., mean over standard deviation) shrinks, but the bias becomes more significant.
Both of these behaviours are as expected.

\begin{figure}
    \centering
    \includegraphics[width=\linewidth]{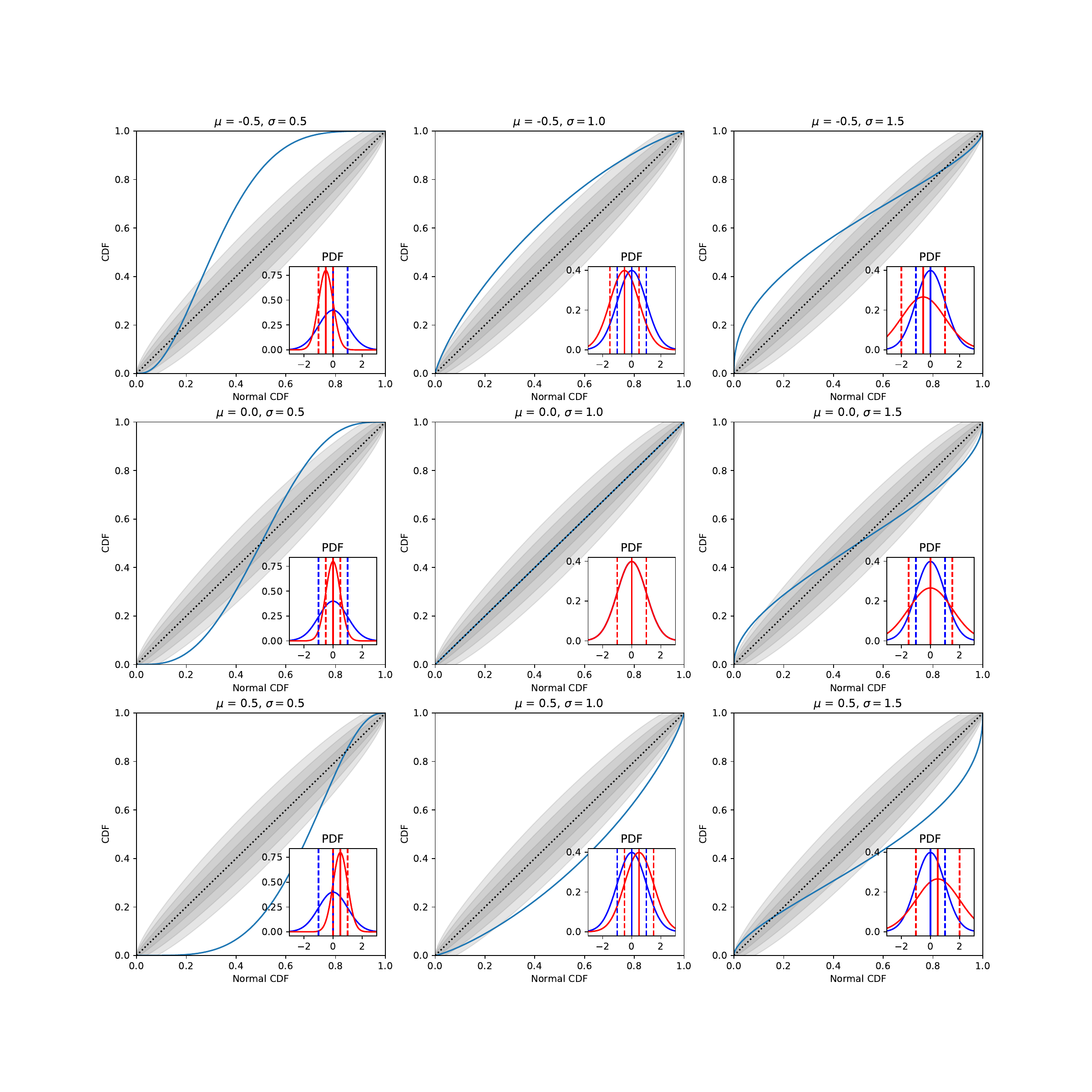}
    \caption{``Dictionary'' for interpreting P-P plots. 
    The blue probability distribution in the inset is the true distribution; the red probability distribution is the recovered distribution.  
    The different rows and columns correspond to different values of the mean (i.e., possible bias) and uncertainty (i.e., possible under or over coverage) of the recovered distributions.
    (Credit: Jessica Lawrence and Kevin Turbang.)} \label{fig:pp-plot-dictionary}
\end{figure}

\begin{figure}[t!]
    \centering
    \includegraphics[width=0.9\linewidth]{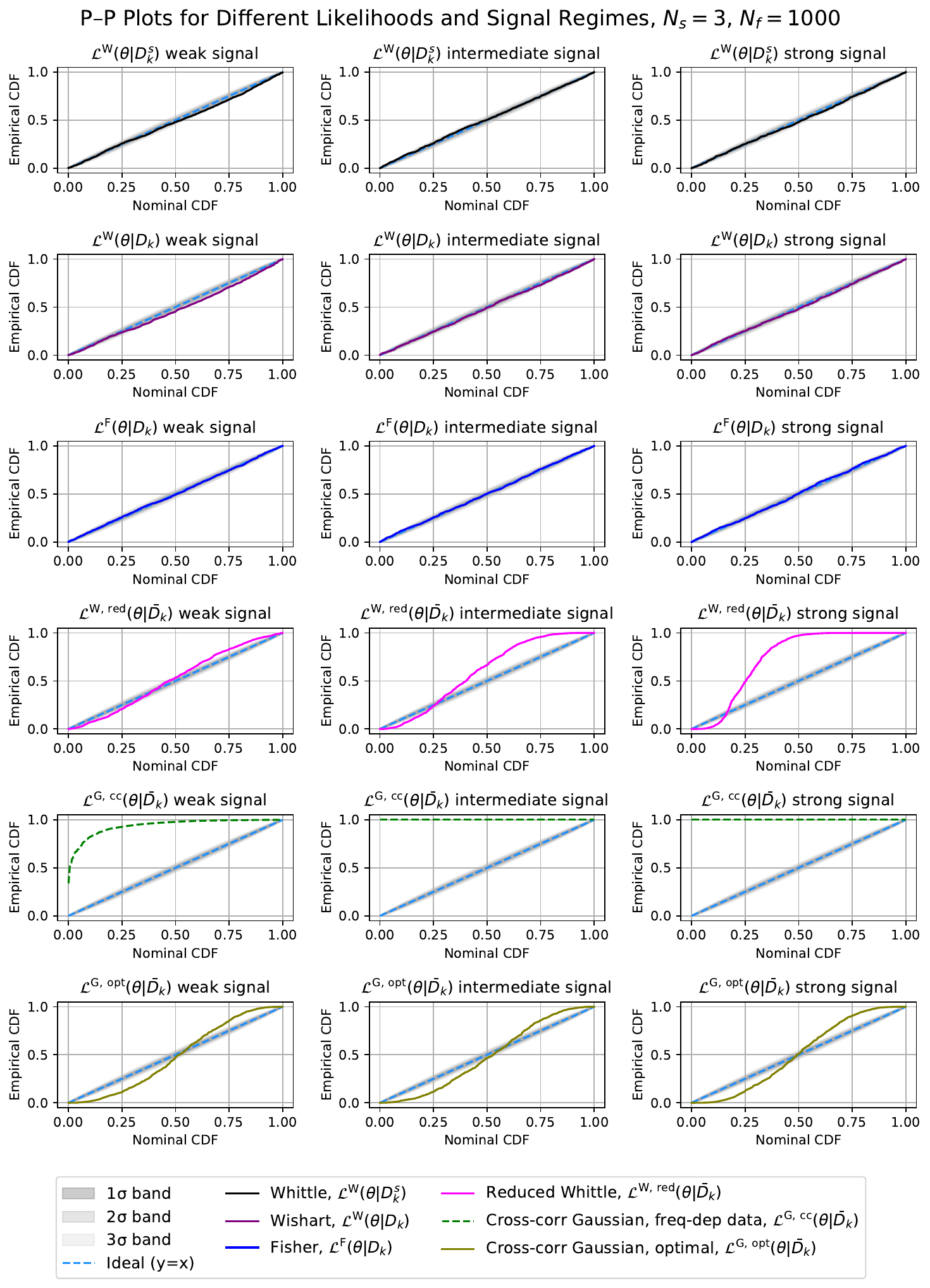}
    \caption{P-P plot for $N_s = 3$ and $N_f = 1000$ for the different likelihoods presented in the paper, for three different signal regimes: ``weak'', i.e., $S_{h}= 0.01$ (left column), ``intermediate'', i.e., $S_{h}= 0.1$ (middle column), ``strong'', i.e., $S_{h}= 1$ (right column). }
    \label{fig:pp-plot1}
\end{figure}

\begin{figure}[t!]
    \centering
    \includegraphics[width=0.9\linewidth]{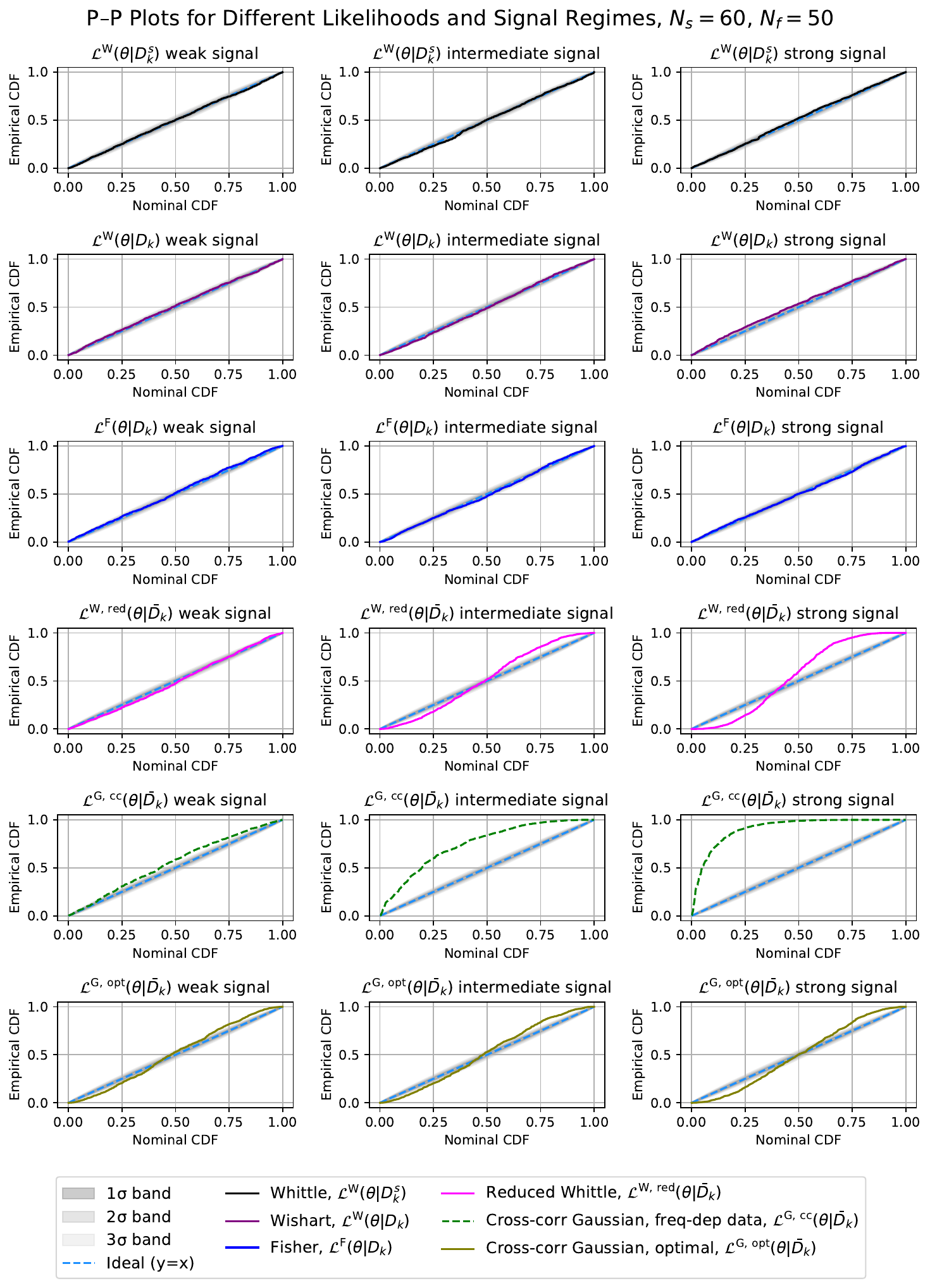}
    \caption{P-P plots as in~\cref{fig:pp-plot1} for $N_s = 60$ and $N_f = 50$ for the different likelihoods presented in the paper, for three different signal regimes: ``weak'', i.e., $S_{h}= 0.01$ (left column), ``intermediate'', i.e., $S_{h}= 0.1$ (middle column), ``strong'', i.e., $S_{h}= 1$ (right column). }
    \label{fig:pp-plot2}
\end{figure}

To aid the interpretation of the probability-probability (P-P) plots, we refer to~\cref{fig:pp-plot-dictionary}, which serves as a practical ``dictionary” for diagnosing common types of patterns. Each panel in~\cref{fig:pp-plot-dictionary} shows the empirical CDF of the posterior quantiles plotted against that of the true CDF (blue line), compared to what you would expect if the two distributions agree (diagonal dotted black line), with shaded bands corresponding to $1\sigma$, $2\sigma$, and $3\sigma$ confidence regions. Insets in each panel display the assumed true (blue) and recovered (red) probability density functions, with vertical dashed lines marking the 68\% confidence interval. The rows and columns correspond to different values of the mean $\mu$ and uncertainty scaling $\sigma$ of the recovered distribution. 
When the P-P plot lies systematically above (below) the diagonal, the inferred mean is biased low (high). Additionally, underestimation of the posterior width leads to an ``S-shaped'' P-P plot, as the injected parameter preferentially falls in the tails of the distribution. On the other hand, an overestimation of the posterior width leads to an ``inverted S'' P-P plot, which indicates the injected parameter preferentially sits in the central quantiles of the distribution.

We proceed by discussing~\cref{fig:pp-plot1} and~\cref{fig:pp-plot2}, which show the P-P plots for the cases where $\Ns=3$, $\Nf=1000$ and $\Ns=60$, $\Nf=50$, respectively, for the ``weak'' ($S_{h}= 0.01$), ``intermediate'' ($S_{h}= 0.1$), and ``strong'' ($S_{h}= 1$) signal regimes. For these plots, we generate 1000 different realizations of the data with the same injected signal value. 

Let us start by discussing~\cref{fig:pp-plot1}. We notice that the P-P plots for the Whittle $\mathcal{L}^{\mathrm{W}} (\theta\vert D^s_k)$, Wishart $\mathcal{L}^{\mathrm{W}} (\theta\vert D_k)$, and Fisher $\mathcal{L}^{\mathrm{F}}(\theta\vert D^s_k)$ likelihoods all lie almost exactly along the diagonal in the weak, intermediate and strong regimes, indicating nearly perfect agreement with the true probability distributions. 
On the other hand, the reduced Whittle likelihood $\mathcal{L}^{\mathrm{W,\,red}}(\theta\vert \bar{D}_k)$ underestimates the variance and is biased low, with both of these deviations becoming more significant as the signal strength increases relative to the noise. The frequency-dependent cross-correlation Gaussian likelihood, $\mathcal{L}^{\mathrm{G,\,cc}}(\theta\vert\bar D_k)$, is also biased low, so much so that for large signals the recovered distribution has minimal overlap with the true distribution.
On the other hand, the ``optimal'' cross-correlation Gaussian  $\mathcal{L}^{\mathrm{G,opt}}(\theta\vert\bar D_k)$ is unbiased, but underestimates the uncertainty in the recovery of the mean.

Figure~\ref{fig:pp-plot2} is similar to \cref{fig:pp-plot1}, but shows what happens when we have more segments ($\Ns=60$ instead of $\Ns=3$) but fewer frequency bins ($\Nf=50$ instead of $\Nf=1000$).
Basically, as $\Ns$ increases, the distributions become more Gaussian and all of the P-P plots become more diagonal, compared to what we saw in \cref{fig:pp-plot1}.
Biases are still present for both the reduced Whittle $\mathcal{L}^{\mathrm{W,\,red}}(\theta\vert \bar{D}_k)$ and frequency‐dependent cross‐correlation Gaussian $\mathcal{L}^{\mathrm{G,\,cc}}(\theta\vert \bar D_k)$ likelihoods (increasing with increasing signal strength), but they are smaller than before because $\Ns$ is larger.
The reduced likelihoods $\mathcal{L}^{\mathrm{W,\,red}}(\theta\vert \bar{D}_k)$, $\mathcal{L}^{\mathrm{G,\,cc}}(\theta\vert \bar D_k)$, and $\mathcal{L}^{\mathrm{G,opt}}(\theta\vert\bar D_k)$
continue to underestimate the variance, since this is a consequence of conditioning the likelihoods on the autopower estimates.
Recall that these estimates are independent of the total number of segments and depend on the number of frequencies only via the smoothing performed when constructing the autopower estimates.

In short, figs.~\ref{fig:pp-plot1} and \ref{fig:pp-plot2} allow for a diagnosis of both the accuracy and precision of the inference procedure using different likelihoods, signal regimes, frequencies, and segments.

\bibliography{main}

\end{document}